\catcode`\@=11

\newif\if@fewtab\@fewtabtrue

{\count255=\time\divide\count255 by 60
\xdef\hourmin{\number\count255}
\multiply\count255 by-60\advance\count255 by\time
\xdef\hourmin{\hourmin:\ifnum\count255<10 0\fi\the\count255}}
\def\ps@draft{\let\@mkboth\@gobbletwo
    \def\@oddhead{}
    \def\@oddfoot
       {\hbox to 7 cm{$\scriptstyle Draft\ version:\ \draftdate$
       \hfil}\hskip -7cm\hfil\rm\thepage \hfil}
    \def\@evenhead{}\let\@evenfoot\@oddfoot}

\def\ceqno{\global\@fewtabfalse
    \ifcase\@eqcnt \def\@tempa{& & &}\or \def\@tempa{& &}
      \or \def\@tempa{&}
      \or\def\@tempa{}\fi\@tempa
{\rm(\theequation)}}

\def\aeqno#1{\global\@fewtabfalse
    \ifcase\@eqcnt \def\@tempa{& & &}\or \def\@tempa{& &}
      \or \def\@tempa{&}
      \or\def\@tempa{}\fi\@tempa
{\rm(\theequation,#1)}}

\def\label#1{\ifnum\draftcontrol=1
 \global\def\draftnote{$\scriptstyle #1$}\fi
 \@bsphack\if@filesw {\let\thepage\relax
   \def\protect{\noexpand\noexpand\noexpand}%
\xdef\@gtempa{\write\@auxout{\string
      \newlabel{#1}{{\@currentlabel}{\thepage}}}}}\@gtempa
   \if@nobreak \ifvmode\nobreak\fi\fi\fi
  \@esphack}

\def\alabel#1#2{\label{#1}\global\@fewtabfalse
    \ifcase\@eqcnt \def\@tempa{& & &}\or \def\@tempa{& &}
      \or \def\@tempa{&}
      \or\def\@tempa{}\fi\@tempa
{\hbox to 3cm{\phantom{\rm(\theequation,#2)}
\draftnote \hfil}\hskip -3cm {\rm(\theequation,#2)}}}

\def\clabel#1{\label{#1}\global\@fewtabfalse
    \ifcase\@eqcnt \def\@tempa{& & &}\or \def\@tempa{& &}
      \or \def\@tempa{&}
      \or\def\@tempa{}\fi\@tempa
{\hbox to 3cm{\phantom{\rm(\theequation)}
\draftnote \hfil}\hskip -3cm{\rm(\theequation)}}}

\def\eqnarray{\def\draftnote{{}}\global\@fewtabtrue
\stepcounter{equation}\let\@currentlabel=\theequation
\global\@eqnswtrue
\global\@eqcnt\z@\tabskip\@centering\let\\=\@eqncr
$$\halign to \displaywidth\bgroup\@eqnsel\hskip\@centering\@eqcnt\z@
  $\displaystyle\tabskip\z@{##}$&\global\@eqcnt\@ne
  \hskip 1\arraycolsep \hfil${##}$\hfil
  &\global\@eqcnt\tw@ \hskip 1\arraycolsep
$\displaystyle\tabskip\z@{##}$
\hfil  \tabskip\@centering&\global\@eqcnt\thr@@\llap{##}\tabskip\z@
\cr}

\def\endeqnarray{\@@eqncr\egroup
      \global\advance\c@equation\m@ne$$\global\@ignoretrue}

\def\@eqnnum{\hbox to 3cm{\phantom{\rm(\theequation)} \draftnote
                         \hfil}\hskip -3cm {\rm(\theequation)}}

\def\@@eqncr{\let\@tempa\relax
    \ifcase\@eqcnt \def\@tempa{& & &}\or \def\@tempa{& &}
      \or \def\@tempa{&}
      \or\def\@tempa{}
\fi\@tempa
\if@eqnsw
\if@fewtab\@eqnnum\fi
\stepcounter{equation}\fi\global
\@eqnswtrue\global\@eqcnt\z@\global\@fewtabtrue\cr}

\def\draftcite#1{\ifnum\draftcontrol=1#1\else{}\fi}

\def\@lbibitem[#1]#2{\item{}\hskip -3cm \hbox to 2cm
{\hfil$\scriptstyle\draftcite{#2}$}\hskip
1cm[\@biblabel{#1}]\if@filesw
     {\def\protect##1{\string ##1\space}\immediate
      \write\@auxout{\string\bibcite{#2}{#1}}}\fi\ignorespaces}

\def\@bibitem#1{\item\hskip -3cm \hbox to 2cm
{\hfil $\scriptstyle\draftcite{#1}$}\hskip 1cm
\if@filesw \immediate\write\@auxout
       {\string\bibcite{#1}{\the\value{\@listctr}}}\fi\ignorespaces}

\def\nsection#1{\section{#1}\setcounter{equation}{0}}

\font\tendl=msbm10  scaled \magstep1
\font\sevendl=msbm7 scaled \magstep1
\font\fivedl=msbm5 scaled \magstep1
\font\tengl=eufm10  scaled \magstep1
\font\sevengl=eufm7 scaled \magstep1
\font\fivegl=eufm5 scaled \magstep1

\newfam\dlfam \def\dl{\fam\dlfam\tendl} 
\textfont\dlfam=\tendl \scriptfont\dlfam=\sevendl
\scriptscriptfont\dlfam=\fivedl
\newfam\glfam  
\textfont\glfam=\tengl \scriptfont\glfam=\sevengl
\scriptscriptfont\glfam=\fivegl

\def\draftdate{\number\month/\number\day/\number\year\ \ \ \hourmin }

\global\def\draftcontrol{0}
\catcode`\@=12
\def\tilde{\widetilde}
\def\hat{\widehat}
\documentstyle[12pt]{article}

\def\theequation{{\thesection.\arabic{equation}}}

\newcommand{\be}{\begin{eqnarray}}
\newcommand{\en}{\end{eqnarray}\vs 0.5 cm}

\newcommand{\no}{\noindent}
\newcommand{\vs}{\vskip}
\newcommand{\hs}{\hspace}

\newcommand{\un}{\underline}
\newcommand{\NR}{{{\dl R}}}
\newcommand{\NP}{{{\dl P}}}
\newcommand{\NC}{{{\dl C}}}
\newcommand{\NZ}{{{\dl Z}}}
\newcommand{\NH}{{{\dl H}}}
\newcommand{\qq}{\begin{eqnarray}}
\newcommand{\de}{\bar\partial}
\newcommand{\da}{\partial}
\newcommand{\ee}{{\rm e}}

\newcommand{\qqq}{\end{eqnarray}}

\newcommand{\tr}{\hbox{tr}}

\newcommand{\CA}{{\cal A}}

\newcommand{\CF}{{\cal F}}

\newcommand{\CH}{{\cal H}}

\newcommand{\CP}{{\cal P}}

\newcommand{\CT}{{\cal T}}

\newcommand{\s}{\hspace{0.05cm}}
\newcommand{\hf}{{_1\over^2}}
\newcommand{\hslash}{{h\hspace{-0.23cm}^-}}
\newcommand{\Di}{{\slash\hs{-0.21cm}\partial}}
\begin{document}
\
\begin{center}
\vskip 0.9cm

{\Large{\bf{CONFORMAL \s FIELD THEORY}}}
\vs 0.3cm

{\Large{\bf and}}
\vs 0.3cm

{\Large{\bf GEOMETRY of \s\s STRINGS}}
\vs 0.5cm

by
\vs 0.7cm

{\large{J\"{u}rg Fr\"{o}hlich}}
\vs 0.2cm

Institut f\"{u}r Theoretische Physik,
ETH-H\"{o}nggerberg,

8093 Z\"{u}rich, Switzerland

\vs 0.5cm

{\large{Krzysztof Gaw\c{e}dzki}}\footnote{\s extended
version of lectures given by the 2${}^{\rm nd}$
author at the Mathematical Quantum Theory Conference
held at the University of British Columbia, Vancouver, Canada,
from August 4 to 8, 1993}
\vs 0.2cm

C.N.R.S., Institut des Hautes Etudes Scientifiques,

91440 Bures-sur-Yvette, France


\end{center}
\date{ }

\vskip 1.6cm

\begin{abstract}
\noindent
What is quantum geometry? This question is becoming
a popular leitmotiv in theoretical physics
and in mathematics. Conformal field theory
may catch a glimpse of the right answer.
We review global aspects of the geometry
of conformal fields, such as duality and mirror
symmetry, and interpret them within Connes'
non-commutative geometry.

\end{abstract}
\vskip 1.3cm

\nsection{\hspace{-.7cm}.\ \ Introduction}

\ \ \ Geometry has been used as a tool
in classical physics in more interesting ways
than in quantum physics. Analytic mechanics
or Einstein's general relativity
are outstanding examples of classical theories
unseparable from their geometric content. It may
seem then that it is enough to look at the quantized versions
of analytic mechanics and of general relativity in order
to understand how the quantum fluctuations modify geometry.
In quantum mechanics, Poisson brackets become
commutators and one could think that quantum
symplectic geometry is the theory of canonical commutation
relations and of their representations. This
would be, however, too naive
as the examples of difficulties with understanding
quantum counterparts of the classical phenomena
of integrability and chaos show. With general relativity,
the situation is even worse: Einstein's gravity has resisted
numerous attempts aimed at quantizing it. These unsuccessful
efforts have convinced physicists that quantum
Riemannian geometry should be rather different from
the classical one, at least at very short distances.
A possible picture of such a modified geometry emerges
from string theory, the best available candidate
for a consistent model of quantum gravity. The fundamental
idea of string theory is to replace point-like objects
by string-like (or loop-like) ones. This should deform
the geometry based on the notion of a space of points.
Quantum string theory is still rather
poorly understood but may be studied, on the classical
and perturbative level, by means of conformal
(quantum, two-dimensional) field theory (CFT). Roughly
speaking, models of CFT describe classical solutions
of string theory. The perturbative expansion around
the classical solutions is built by considering the CFT models
on two-dimensional space-times of non-trivial topology.
To understand the string geometry on the classical
and perturbative level, one should then understand
the (quantum) geometry of CFT's. Although a fully quantum
string geometry at arbitrarily small scales
may not be accessible to
classical or perturbative analysis, its behavior
in the large should be captured by the classical approximation.
Below we shall concentrate on phenomena which distinguish
the geometry in the large of CFT's from the conventional
Riemann-Einstein geometry.
\vs 0.3cm

How should a physicist think about Riemannian geometry?
As in other cases, she or he should extract general
concepts from observable quantities. The trajectories
of small bodies (test particles) are observable
and, in general relativity, they are the time-like
geodesics. The latter, when parametrized by the proper
time, encode the complete information about the pseudo-Riemannian
metric. The Lorentzian signature of the relevant geometry
is, of course, the basic physical fact but, below, we shall
limit ourselves very early to the easier case of euclidean
signature. Technically, we shall use the tools
of non-commutative geometry which were developed
in the euclidean context. Understanding why the signature of
the effective space-time is indefinite remains one of the
principal open problems of quantum gravity
and we shall have little to add here besides pointing out
that the development of a Lorentzian non-commutative geometry
seems to be, from this perspective, a natural task.
\vs 0.3cm

Let \s$M\s$ be a (pseudo-)Riemannian
manifold with metric \s$\gamma=\gamma_{\mu\nu}\s dx^\mu dx^\nu\s$.
One usually takes the length
\qq
\int ds=\int\s[\gamma_{\mu
\nu}(x)\s
{dx^\mu\over{d\tau}}\s{dx^\nu\over{d\tau}}]^{1/2}\s d\tau
\label{length}
\qqq
of the trajectory \s$\tau\mapsto x(\tau)\in M\s$ as the action
functional for the test particle (with unit mass).
The extrema of such an action
are arbitrarily parametrized geodesics. It will be more convenient
to use
\qq
S(x(\cdot))\s=\s{_1\over^2}\int
\gamma_{\mu\nu}(x)\s{_{dx^\mu}\over^{d\tau}}\s
{_{dx^\nu}\over^{d\tau}}\s d\tau
\label{gf}
\qqq
as the action, with the stationary points given
by the geodesics parametrized by a constant times the length.
If, following
Polyakov's approach \cite{Polya}, we couple the
latter action to the world-line metric \s$h(\tau)d\tau^2\s$,
\s replacing it by
\qq
S(x(\cdot),h(\cdot))\s=\s{_1\over^2}\int
\gamma_{\mu\nu}(x)\s{_{dx^\mu}\over^{d\tau}}\s
{_{dx^\nu}\over^{d\tau}}\s h^{-1/2} d\tau
\s+\s{_1\over^2}\int h^{1/2}d\tau\ ,
\label{action2}
\qqq
then we can recover the length of the trajectory
by minimizing (\ref{action2}) over \s$h(\cdot)\s$: \s
classically, the two
actions are equivalent. For \s$h(\tau)\equiv 1\s$,
\s the action (\ref{action2}) reduces to
(\ref{gf}) which describes a mechanical system called
the ``geodesic flow'' on \s$M\s$. Setting \s$h(\tau)\s$ to
\s$1\s$ may be viewed as a gauge choice fixing the
reparametrization invariance of the action (\ref{length}).
\vs 0.3cm

It is easy to quantize the geodesic motion of a test
particle on a Riemannian manifold\footnote{\s on pseudo-Riemannian
manifolds a fully consistent quantization requires, however,
a multi-particle approach}. The Hilbert space
of states \s$\NH\s$ may be taken as the space
\s$L^2(M,dv_\gamma)\s$ of functions
on \s$M\s$ square-integrable with respect to
the Riemannian volume
\s$dv_\gamma\s$. \s Physically, this is the space of functions on
the configuration space of the particle.
As the (positive) Hamiltonian \s$\CH\s$ governing the quantum
evolution, we may take \s$-{_1\over^2}\Delta_\gamma\s$, \s where
\s$\Delta_\gamma\s$ is the Laplace-Beltrami operator
on \s$M\s$. \s In a first step towards understanding
what  quantum Riemannian geometry might be, one may
try to reformulate standard Riemannian geometry
using the fundamental notions of quantum mechanics:
those of a Hilbert space of states and of an algebra
of observables. Somewhat surprisingly, the
exercise has proven to be rich in consequences. It has
led A. Connes to the development of
non-commutative geometry \cite{Connes1} providing an
extension of geometry to situations very far from its
original context. One such situation, which
required an extension of Connes' geometry to an
infinite-dimensional setup, was the analysis of models
of two-dimensional massive quantum supersymmetric field
theory \cite{JafLes}\cite{JafLesOs}.
Those models, although not exactly
solvable, could be controlled by the analytic methods of
constructive field theory \cite{GJ}. We shall argue below
that non-commutative geometry (already in its finite-dimensional
version) may provide tools to study the deformed
geometry of the exactly solvable (massless) models of CFT.
\vs 0.3cm

The abstract pair \s$({\NH},-\hf\Delta_\gamma)\s$
does not, in general,
determine the geometry of \s$M\s$ \cite{Milnor}: one cannot
hear the shape of the drum \cite{Kac}. We need more
structure. Such additional structure
is provided by the algebra \s$\CA\s$ of
observables measuring the position of the test particle,
realized as the algebra of multiplication operators
by (say, smooth,
bounded) functions on \s$M\s$. The manifold \s$M\s$ may
be reconstructed from \s$\CA\s$. The points of \s$M\s$
may be identified with the characters of (the norm closure of)
\s$\CA\s$. The differentiable structure of \s$M\s$ is then
determined since we know the smooth functions on it.
That the abstract
triple \s$({\NH},-\hf\Delta_\gamma,\CA)\s$
determines also the Riemannian metric
is implied\footnote{\s we thank J. Derezi\'{n}ski for this observation}
by a slight modification of A. Connes' argument
\cite{ConnesLott} which goes as follows:
the geodesic distance between points \s$x\s$ and \s$y\s$ of
\s$M\s$ (which are multiplicative linear functionals on
\s$\CA\s$) \s is given by
\qq
d_\gamma(x,y)=\rm{sup}\s|f(x)-f(y)|\ ,
\qqq
where the supremum is taken over smooth bounded functions
\s$f\s$ s.t.
\qq
{_1\over^2}(\Delta_\gamma f^2+f^2\Delta_\gamma)-f\Delta_\gamma f\ ,
\qqq
which is the multiplication operator
by \s$\langle df,df\rangle_\gamma\equiv
\gamma^{\mu\nu}(\da_\mu f)(\da_\nu f)\s$, \s
has norm \s$\leq\s1\s$.\s\s
Similarly, much of classical Riemannian geometry
may be rewritten in terms of the quantum mechanics of a particle
moving on \s$M\s$ in a way that uses little of the particular
properties of \s$({\NH},-\hf\Delta_\gamma,A)\s$. \s Connes' idea
was then
to use triples like \s$({\NH},\CH,A)\s$, \s where \s$\CH\s$
is a self-adjoint operator on \s$\NH\s$ and \s$\CA\s$ a
$*$-algebra of bounded operators on \s$\NH\s$ (in general
non-commutative) as the starting point for non-commutative
geometry\footnote{\s in fact, he uses the Dirac operator instead
of the Laplacian, an important refinement which we shall discuss
below}.
\vs 0.3cm

Let us now proceed from mechanics to 1+1-dimensional
field theory, with the 1-dimensional space compactified
to a circle. On the classical level, a large family
of such theories is given by the so called ``sigma models''.
They are obtained by considering
fields \s$x(\sigma,\tau)\s$ with values in the Riemannian
manifold \s$M\s$, \s with the action functional
\qq
S(x(\cdot,\cdot))\s=\s{_1\over^{4\pi}}\int\gamma_{\mu\nu}(x)
\s(\da_\tau x^\mu\s\da_\tau x^\nu-\da_\sigma
x^\mu\s\da_\sigma x^\nu)\s d\sigma\s
d\tau\ .\label{action3}
\qqq
We assume periodicity of \s$x(\cdot,
\cdot)\s$ in the space variable \s$\sigma\s$ with period
\s$2\pi\s$. \s The \s$\sigma\s$ integration in (\ref{action3})
\s is restricted to the interval \s$[0,2\pi[\s$.
It is sometimes useful to consider more general sigma models,
with the action modified by the addition of the term
\qq
S'(x(\cdot,\cdot))\s=\s{_1\over^{2\pi}}\int
\beta_{\mu\nu}(x)\s\da_\sigma x^\mu\s\da_\tau
x^\nu\s d\sigma\s d\tau\ ,
\label{action4}
\qqq
where \s$\beta_{\mu\nu}dx^\mu\wedge dx^\nu\equiv\beta\s$
is a 2-form on \s$M\s$. \s
The classical solutions of the stationarity equations for
the sigma model action are
parametrized, at least for small times, by
the Cauchy data
\s$x(\cdot,0)\s$ and \s$\da_\tau x(\cdot,0)\s$, \s
so that the space \s$LM\s$ of (smooth) loops
\s$\sigma\mapsto x(\sigma,0)\s$
in \s$M\s$ plays the role of the configuration space of
the model. \s$LM\s$ is, itself, a Riemannian
manifold (in the Fr\'{e}chet sense) with the metric
induced from that of \s$M\s$:
$$\|\delta x\|^2\s=\s{_1\over^{2\pi}}\int\gamma_{\mu\nu}
(x(\sigma))\s\s\delta x^\mu(\sigma)\s
\delta x^\nu(\sigma)\s d\sigma\ .$$
The replacement of \s$M\s$ by \s$LM\s$ may seem
to be the essential step of string geometry.
Notice, however, that the sigma model action
(\ref{action3}) differs from the one for the
geodesic motion on \s$LM\s$ by a ``potential''
type term with spatial derivatives of \s$x^\mu\s$.
Besides, it is not the classical
1+1-dimensional field theory but the corresponding
quantum theory
which describes the classical (and perturbative)
level of string theory and, consequently, we should
describe the geometry of the quantum sigma models, not that
of the classical ones. The latter is fairly directly related to
geometry of the loop space \s$LM\s$. The quantum sigma models,
however, require a renormalization of the target space geometry.
If one attempts to construct them in a perturbation
expansion in powers of the Planck constant \s$\hslash\s$,
\s each order introduces (infinite) counterterms modifying
the initial Riemannian metric of \s$M\s$ \cite{Friedan}.
As a result of the renormalization, the direct relation
between quantum theory and geometry of the
target manifold \s$M\s$ or of its loop space \s$LM\s$
is blurred. This is a new phenomenon since, as we have
seen, quantum mechanics of a particle on a Riemannian
manifold fully encodes the geometry of the manifold.
To the first order in \s$\hslash\s$, \s there is no metric
renormalization and a sigma model
defines, in this approximation, a CFT if and only if
the target metric is Ricci flat, i.e.
if it solves the Einstein equations \cite{Friedan}.
Clearly, we may view the Riemann-Einstein geometry as a
limiting case of the stringy
one.
\vs 0.3cm

Renormalization renders a rigorous
construction of generic sigma models
very difficult. There exists, however, a big
pool of exactly solvable 1+1-dimensional quantum field
theory models. Among those, there is a rich family of CFT's
where we know exact expressions not
only for the energy eigenvalues but also for
Green's functions encoding the operator product of
field operators. With any model of quantum field
theory, we may associate a triple \s$({\NH},\CH,A)\s$, \s
where \s${\NH}\s$ is the Hilbert space of states,
\s$\CH\s$ is the Hamiltonian
and \s$\CA\s$ is the (non-commutative) algebra generated by the
field operators.
We may think of this triple as encoding the effective
quantum geometry of the space of field configurations
or of its cotangent bundle. For the sigma models,
this effective geometry summarizes the
deformation, due to the renormalization effects, of the
infinite-dimensional geometry of the loop space \s$LM\s$.
The presentation of a quantum model by the triple
\s$({\NH},\CH,A)\s$, selecting the Hamiltonian,
is more natural in mechanics than in field theory, where it requires
a choice of the time direction on the world-sheet. There, it
would be more appropriate to specify the energy-momentum
vector \s$(\CH,\CP)\s$ or, in CFT, the whole set
of generators of the conformal algebra or of its extensions.
In fact, the most convenient algebraic setup for describing
non-commutative geometry of quantum fields still remains to
be found. Below, we shall concentrate on
geometry of the low energy modes of quantum field theory
where quantum mechanical description is quite sufficient.

\vs 0.3cm

On the classical level, it is easy to recover
the geodesic motion of the particle on \s$M\s$ from
the sigma model: it is enough to consider
1+1-dimensional fields which do not depend
on the space variable \s$\sigma\s$. \s Notice that,
for such fields, the action (\ref{action3}) reduces to
(\ref{gf}) and (\ref{action4}) disappears.
The space of \s$\sigma$-independent fields realizes
an embedding of \s$M\s$ into the subspace of
elements of \s$LM\s$ invariant under reparametrizations.
Although it is difficult to construct sigma models
by quantizing the classical theory which
has \s$M\s$ as the target, one may ask if it is possible
to identify the effective target starting from a
1+1-dimensional quantum field model. Below, we shall show
that, indeed, this can be done for the CFT models by using there
infinite-dimensional symmetries like those given by
the Virasoro algebra (essentially the algebra of
infinitesimal reparametrizations of the circle), the
current algebras or the supersymmetric extensions of those.
With the use of such a symmetry algebra, one may associate
to a CFT model a triple \s$({\NH}_0,\CH_0,\CA_0)\s$
describing the effective quantum geometry of
zero modes of the string. The latter dominate the low energy regime
where the internal motion of the string may be neglected.
Usually the algebra \s$\CA_0\s$ will be non-commutative,
the commutativity being restored only in a semiclassical
limit. Sometimes several distinct families of triples
\s$({\NH}_0,\CH_0,\CA_0)\s$, \s yielding
the effective low energy description in different
limiting regimes, may be associated
to a given family of conformal models. Consequently,
one CFT may correspond to different (in general non
commutative) effective targets.
As we shall see, this is the essence of the duality \cite{GSB}
and mirror symmetry \cite{GreenPless}\cite{CLSch}
phenomena which are among the most
interesting novel features of string geometry in the large.
\vs 0.3cm

Certainly, string geometry is still in an early stage
of development. What we have at our disposal are numerous
examples of CFT models with rich infinite-dimensional
symmetries. They may be thought of as symmetric
spaces of string geometry. Their study is a stringy version
of Klein's ``Erlanger Programm''. What is still largely
missing is a stringy version of Riemann's approach to
geometry. Such an approach should probably pass through
string field theory which has common points
with Connes' non-commutative geometry, as Witten's
open string field theory \cite{Wittenopen} has shown.
What is described below, is a more timid
attempt to develop one limited aspect of
string geometry: that of the effective metric geometry of \s
low energy states. We point out that Connes'
theory provides useful tools also for this more limited
and more phenomenological approach.
There are other geometric aspects of the low energy
string like those involving dilatonic and Yang-Mills geometry.
They require studying more general
sigma models whose examples we shall encounter examining
coset conformal theories and their supersymmetric
versions. The non-metric aspects of string geometry
should be related to the non-commutative $K$-theory
and certainly deserve to be studied.
One of the shortcomings of our approach (present already
in the treatment of the relativistic quantum mechanics)
is that we work with the world-sheet (world-line)
metric fixed ignoring the effects of fluctuations
of the latter. In this way, what we study is more
geometry of the conformal fields than that of the string.
This is not necessarily a drawback since
CFT geometry may be more appropriate for the classical description
of non-topological phases of the string. Nevertheless,
a deeper understanding of string
geometry should take into account the fluctuations of
the world-sheet backgrounds which, treated with BRST
techniques, play an important role in string field theory.
\vs 0.4cm

The following is the plan of the present expos\'{e}:
\vs 0.3cm

\no In Section 2, we shall describe the sigma models
with a circle as the target. They are essentially
versions of free 1+1-dimensional fields. These models
illustrate the duality phenomenon responsible
for the appearance of a fundamental length scale in
string geometry. Their slight generalization
with complex tori as the targets allows also to exhibit
the simplest instance of the mirror symmetry \cite{Yau}.
\vs 0.3cm

\no Section 3 will be a guided tour through the
factory of symmetric models of CFT. The basic raw
material for the production of those
models is the Wess-Zumino-Witten
(WZW) 1+1-dimensional
field theory which, when processed by a gauging
machine, gives a rich family of ``coset models''
of CFT. We shall briefly explain how the
gauging machine works.
\vs 0.3cm

\no Section 4 will be devoted to geometry
of the supersymmetric CFT models. We shall present the
supersymmetric version of the WZW
theory and of the coset models.
\vs 0.3cm

Finally, in Section 5, we shall sketch
the relation between CFT models with N=2
supersymmetry and Calabi-Yau geometry. This
relation lies at the core of the mirror symmetry phenomenon
for which non-commutative geometry provides a
natural framework.
\vs 0.8cm

\no{\bf{Acknowledgements.}} \ We thank A. Connes for supporting
us in our struggle to learn some non-commutative
geometry and C. Voisin for teaching us basic complex
geometry. In working on the program described in these
notes we profited from numerous discussions with
A. H. Chamseddine and G. Felder. J.F. thanks the I.H.E.S.
and K.G. the Forschungsinstitut f\"{u}r Mathematik of E.T.H.
for hospitality
which made this collaboration possible. We are also
greatful to J.Feldman, R. Froese and L. Rosen, the organizers
of the Vancouver meeting, for the possibility to present
our ideas to a mathematical physics audience.
\vskip 0.8cm

\nsection{\hspace{-.7cm}.\ \ Toroidal geometry}
\vs 0.2cm
\no{\bf{2.1.\s\ \ \s$S^1\s$ target and duality}}
\vs 0.4cm

The simplest sigma model is obtained by taking the circle
\s$\NR/2\pi\NZ\s\equiv S^1$ as the target. We shall
use the angle variable \s$x\s$ as the coordinate of
the circle. It may be assumed that the
Riemannian metric of \s$S^1\s$ takes the form
\s$\gamma=r^2 dx^2\s$,
\s since the radius \s$r\s$ is the only metric invariant.
In other words, \s$r>0\s$ parametrizes the half line
of Riemannian circles. Encoding the geometry
of the circle in quantum mechanics of a particle
moving on it gives rise to the triple
\qq
(\s L^2(S^1\hs{0.03cm},\s{_{dx}\over^{\sqrt{2\pi}}})\s,\ -\hf\s
r^{-2}{_{d^2}\over^{dx^2}}\s,\ C^\infty(S^1)\s)\ .
\qqq
Functions \s$x\mapsto\ee^{inx}\s$ are the eigenfunctions
of the Laplacian corresponding to the eigenvalues
\s$r^{-2}n^2\s$.
The action of a sigma model with the \s$S^1\s$ target is
\qq
S(x(\cdot,\cdot))\s=\s{_{r^2}\over^{4\pi}}\int((\da_\tau x)^2-
(\da_\sigma x)^2)\s d\sigma\s
d\tau\ .
\qqq
Its stationary points satisfy the wave equation
\qq
(\da_\tau^2-\da_\sigma^2)\s x\s=\s 0
\qqq
with the general solution
\qq
x(\sigma,\tau)\s=\s x^0+r^{-2}p\tau+w\sigma+\sum
\limits_{n\not=0}{_1\over^{\sqrt{2}\s i\s rn}}\s\alpha_{n}^+
\s\ee^{in\s(\sigma+\tau)}
+\sum
\limits_{n\not=0}{_1\over^{\sqrt{2}\s i\s rn}}\s
\alpha_{n}^-\s\ee^{-in\s(\sigma
-\tau)}
\ .
\label{gs}
\qqq
$w\in\NZ\s$ is the winding number, a homotopy invariant of the
solution. It labels the connected components of the space
of solutions. The canonical Poisson brackets
\qq
\{\s x(\sigma,0)\s,\s\da_\tau x(\sigma',0)\s\}\s=\s
2\pi r^{-2}\s\delta(\sigma-\sigma')
\qqq
translate to
\qq
\{\s x^0\s,\s p\s\}=1\s,\ \ \
\{\s\alpha_{n}^+\s,\s\alpha_{m}^+\s\}=
in\s\delta_{n+m,0}\s,\ \ \ \{\s\alpha_{n}^-\s,
\s\alpha_{m}^-\s\}=
in\s\delta_{n+m,0}
\qqq
with all the other Poisson brackets vanishing.
Classically, the model posseses the conformal
symmetry acting by reparametrizations
of \s$\tau\pm\sigma\s$ and the symmetry
\qq
x(\sigma,\tau)\s\mapsto\s x(\sigma,\tau)\pm\delta^\pm
x(\tau\pm\sigma)\ ,
\label{symm}
\qqq
where \s$\delta^\pm x(\cdot)\s$ are arbitrary (periodic) functions.
\vs 0.3cm

Since we are essentially in the free field case,
quantization of the model, replacing
Poisson brackets by \s$i\s$ times commutators,
is standard. The Hilbert space of states of the quantized
system is
\qq
{\NH}\s=\s L^2(S^1\hs{0.03cm},{_{dx}
\over^{\sqrt{2\pi}}})^\NZ\otimes\CF^+
\otimes\CF^-\ .
\qqq
Above, the exponent \s$\NZ\s$ indicates the infinite direct sum
of \s$L^2\s$ spaces with the components labeled by the winding
numbers \s$w\in\NZ\s$. Periodic functions of \s$x^0\s$ act
by multiplication and \s$p\s$ as \s$i{{d}
\over{dx}}\s$ in each \s$L^2\s$ component.
We may span \s$L^2(S^1\hs{0.03cm},{{dx}
\over{\sqrt{2\pi}}})^\NZ\s$ by
the eigenvectors \s$|p_w,w\rangle=
\ee^{-ip_wx}\s,\ p_w\in\NZ\s,$
\s of \s$i{{d}\over{dx}}\s$ in each component of
the direct sum. We shall also label states \s$|p_w,w\rangle\s$
as \s$|p^+;p^-\rangle\s$, \s where the left-right momenta
\qq
p^\pm={_1\over^{\sqrt{2}}}(r^{-1}p_w\pm rw)\ .
\label{lrmom}
\qqq
The set \s\s$\{\s (p^+,p^-)\s\}\s\s$ provided with the
quadratic form \s$(p^+)^2-(p^-)^2=2p_ww\s$ forms an even
self-dual Lorentzian lattice.
$\CF^\pm\s$ are two copies of the
bosonic Fock space built on the vacuum state \s$|0\rangle\s$
annihilated by \s$\alpha_n\s,\ n>0\s,$ \s via the action of
operators \s$\alpha_n\s,\ n<0\s$. \s$[\alpha_n,\alpha_m]=n
\s\delta_{n+m,0}\s$ and \s$\alpha_n^*=\alpha_{-n}\s$.
We shall consider \s$L^2(S^1)^\NZ\s$ as embedded into
\s${\NH}\s$ by \s$|p^+;p^-\rangle\s\mapsto\s|p^+;p^-\rangle
\otimes|0\rangle_+\otimes|0\rangle_-\s\s$.
\vs 0.3cm

Let us introduce chiral (multivalued) quantum fields
\s$X_\pm\s$,
\qq
X_\pm(\tau\pm\sigma)\s=\s X_\pm^0 +
p^\pm(\tau\pm\sigma)+\sum\limits_{n\not=0}
{_1\over^{in}}\alpha^\pm_n\ee^{in\s(\tau\pm\sigma)}\ ,
\qqq
where \s$X_\pm^0={1\over i}{d\over{dp^\pm}}\s$
acts in the extended space with arbitrary momenta.
On the quantum level, symmetry (\ref{symm}) is
generated by a commuting pair of \s$u(1)\s$
current algebras with the currents
\qq
&j^\pm(\tau\pm\sigma)=\da_\sigma\s X_\pm(\tau\pm\sigma)
\equiv\sum\limits_{n=-\infty}^\infty
j^\pm_n\s\ee^{in\hs{0.03cm}(\tau\pm\sigma)}&\ ,\\
&[j_n\s,\s j_m]=n\s\delta_{n+m,0}\ \ &
\qqq
($j^\pm_0=\pm p^\pm\s,\ \s j_n^\pm=\pm\alpha^\pm_n\s$ for
\s$n\not=0\s$)\s. \s The Hilbert space \s$\NH\s$ is a direct sum
of the irreducible (highest weight) representations
of \s$j^\pm\s$  acting in \s$|p^+;p^-\rangle\otimes\CF^
+\otimes\CF^-\s$, \s labeled
by \s$u(1)\s$ charges \s$\pm p^\pm\s$. \s
The classical conformal symmetry
also carries over to the quantum level. It is
represented there by a commuting pair of Virasoro algebras
given by the Sugawara construction \cite{Suga}:
\qq
L^\pm_n=\hf\sum\limits_{m=-\infty}^\infty:j^\pm_m\s j^\pm_{n-m}:\ ,
\qqq
where the Wick ordering puts operators \s$\alpha^\pm_n\s$
with \s$n<0\s$ to the left of the ones with \s$n>0\s$.
\qq
[L^\pm_n\s,\s L^\pm_m]=(n-m)L^\pm_{n+m}+{_c\over^{12}}
(n^3-n)\s\delta_{n+m,0}\ ,
\qqq
with the Virasoro central charge \s$c=1\s$.
The Hamiltonian of the model is \s$\CH=L^+_0+L^-_0-{1\over 12}\s$
and the momentum operator generating the space
translations is \s$\CP=L^+_0-L^-_0\s$.
\qq
\CH\s=\s\hf\s(r^{-2}p^2+r^2w^2)\s+\s\sum
\limits_{n>0}\alpha_{-n}^+\s\alpha_{n}^++\s\sum
\limits_{n>0}\alpha_{-n}^-\s\alpha_{n}^-\s-\s{_1\over^{12}}
\qqq
and $|{\rm vac}\rangle\equiv|0;0\rangle\s$ is its
unique eigenstate (with eigenvalue \s$-{1\over 12}\s$)
at the bottom of the spectrum (the vacuum).
\vs 0.3cm

The basic quantum fields of our model are given
by ordered exponentials of fields \s$X_\pm\s$
\qq
\tilde V_{q^+q^-}(z_\pm)\s\equiv\s\s
:\ee^{\s -i\s q^+X_+(\tau+\sigma)\s-\s i\s q^-X_-(\tau-\sigma)}:\ ,
\qqq
where \s$z_\pm\equiv\ee^{-i\s(\tau\pm\sigma)}\s$ and
the Wick ordering puts also the operators \s$X_\pm^0\s$
to the left of \s$p^\pm\s$.
When
\qq
q^\pm\s=\s{_1\over^{\sqrt{2}}}(r^{-1}n\pm rm)
\qqq
for \s$n,\s m\in\NZ\s$, \s then the above
operators\footnote{\s or rather
their smeared versions} map (a dense subset of) our
Hilbert space into itself. Corrected by the cocycles
\s$c_{q^+q^-}(p^+,p^-)\equiv(-1)^{(q^++q^-)(p^+-p^-)/2}=(-1)^{nw}\s$
they give mutually local vertex operators
\qq
V^{q^+q^-}(z_\pm)=\tilde V^{q^+q^-}(z_\pm)\s\s c_{q^+q^-}\ .
\qqq
$V_{q^+q^-}\s$ are primary fields
of charges \s$\pm q^\pm\s$ for the \s$u(1)\s$
currents \s$j^\pm\s$, \s i.e. they satisfy the following
commutation relations:
\qq
[\s j^\pm_n\s\s,\s\s V_{q^+q^-}(z_\pm)\s]\s
=\s \pm q^\pm\s\s z_\pm^n\s\s V_{q^+q^-}(z_\pm)\s\s
\label{pf}
\qqq
which say that \s$V_{q^+q^-}\s$ transform
as local fields of \s$u(1)\s$ charges \s$\pm q^\pm\s$ under
the local gauge transformations induced by \s$j^\pm\s$.
Fields \s$V_{q^+q^-}\s$ are also primary fields of the
Virasoro algebras:
\qq
[\s L^\pm_n\s,\s V_{q^+q^-}(z_\pm)\s]\s=\s
\Delta_{q^\pm}^\pm(n+1)\s z_\pm^{n}\s V_{q^+q^-}(z_\pm)\s+
\s z_\pm^{n+1}\s\da_{z_\pm}V_{q^+q^-}(z_\pm)
\label{Virpf}
\qqq
i.e. they transform under the conformal transformations
as tensors with weights \s$\Delta^\pm_{q^\pm}\equiv
{1\over 2}(q^\pm)^2\s$.
As usual in CFT, the primary fields may be labeled
by specific vectors
\qq
\lim_{z_\pm\rightarrow 0}\s V_{q^+q^-}
(z_\pm)\s|{\rm vac}\rangle\s=\s|q^+;q^-\rangle
\label{pf-v}
\qqq
in the Hilbert space (above, the values of \s$z_\pm\s$
are extended to the complex domain by
analytic continuation).
\vs 0.4cm

Defining the (non-commutative) algebra \s$\CA\s$
to be generated by the vertex
operators, we obtain a triple
\qq
({\NH},\CH,\CA)_r
\label{TR}
\qqq
associated to the sigma model with the circle
of radius \s$r\s$ as the target.
It may be viewed as encoding
geometry of the sigma model.
\vs 0.3cm

The following simple observation has
deep consequences. \s${\tilde j}^\pm\equiv
\pm j^\pm\s$ is an equally
good commuting pair of \s$u(1)\s$ currents in the sigma
model with the \s$S^1\s$ target
as \s$j^\pm\s$. Now, the representation content of the
radius \s$r\s$ model with respect to \s${\tilde j}^\pm\s$
is the same as of the radius \s$r^{-1}\s$ model with
respect to \s$j^\pm\s$. The Hilbert
spaces of states for radii \s$r\s$ and \s$r^{-1}\s$ may
then be identified by the duality transformation
sending \s$|p^+,p^-\rangle\s$ to
\s$(-1)^{p_ww}|p^+,-p^-\rangle\s$ (i.e., modulo signs,
interchanging the roles
of the momentum and of the winding number) and
\s$\alpha^\pm_n\s$ to \s$\pm\alpha^\pm\s$. Under this
identification, the Hamiltonians and the algebras
\s$\CA\s$ (as well as the entire CFT's) coincide:
\qq
({\NH},\CH,\CA)_r\s\cong\s({\NH},\CH,\CA)_{r^{-1}}\ .
\qqq
This phenomenon distinguishes the stringy
geometry of circles from their point-set Riemannian
geometry. While the space
of Riemannian circles was the half line parametrized by
the radius \s$r\s$, that of the sigma models with circle
targets is an orbifold of the latter obtained by the
identification of \s$r\s$ with \s$r^{-1}\s$.
\vs 0.3cm

The above is an example of a more general duality
phenomenon responsible for the appearance of a fundamental
length scale in string theory. This is certainly one of the most
promising features of the latter. In order to understand
how such a length scale arises, let us think of how one probes
the effective space(-time) geometry in string theory.
As mentioned before, this should be done by looking at the
low energy states\footnote{\s with the energy measured from
the bottom of the spectrum} of the string in which
the stringy internal modes are not excited
and one effectively sees quantum mechanics of point-like
objects. Let us suppose that
the string vacuum is described by a sigma model with the \s$S^1\s$
target. The oscillatory modes of the string created by
operators \s$\alpha^\pm_n\s,\ n<0\s$ have energies quantized
to integer values, so we should look at the states with
energies $\ll 1\s$. \s If the radius \s$r\s$ of the circle
is much bigger than \s$1\s$ (or than the Planck length
in dimensional units) then the low-energy spectrum of the sigma
model is given by the states \s$|p_0,0\rangle\s$,
\s$p_0\in\NZ\s$, \s describing the baricentric
degree of freedom \s$x^0\s$. We effectively obtain
the quantum mechanics of a particle moving on the circle
of radius \s$r\s$. If \s$r\s$ is much smaller
than \s$1\s$ then the low-energy spectrum of the sigma model
is given by the states \s$|0,w\rangle\s$ corresponding to the
winding modes of the string. But this is exactly like
the spectrum in the quantum mechanics of a particle moving
on the circle of radius \s$r^{-1}\s$. \s As a result, we
never see circles of small radii as effective geometries!
Notice that this is a quantum phenomenon as is signaled
by the presence of \s$\hslash\s$ in the expression
for the Planck length.
\vs 0.3cm

We may formalize this observation in the following way.
Notice that the classical states invariant under
(\ref{symm}) with \s$\int\delta^\pm x=0\s$
depend only on the zero mode variables \s$x_0,\s p\s$
and \s$w\s$. On the quantum level, we may then
define the zero mode states as those annihilated by the
positive frequency part of the $u(1)$
currents
\qq
{\NH}_0\s=\s\{\ |\phi\rangle\in{\NH}\ \s\s|\ \s\s
j^\pm_n|\phi\rangle=0
\ \s{\rm for}\s\ n>0\ \}\ .
\qqq
${\NH}_0\s\s$ is the subspace of the highest weight vectors
for the \s$u(1)\times u(1)\s$ current algebra. It is
spanned by vectors
\s$|p^+;p^-\rangle\s$ with \s$p^\pm\s$ as in (\ref{lrmom}).
The Hamiltonian \s$\CH\s$ leaves \s${\NH}_0\s$ invariant:
\qq
(\CH+{_1\over^{12}})\s|p^+;p^-\rangle\s
=\s\hf(r^{-2}p_w^2+r^2w^2)|p^+;p^-\rangle\ .
\qqq
Notice that not all states in
\s$\NH_0\s$ have energy \s$\ll 1\s$. \s In the semiclassical
regime when \s$r\gg 1\s$
this is possible only for states with winding number
\s$w=0\s$.
We shall then define a subspace
\qq
{\NH}'_0&=&\{\ |\phi\rangle\in{\NH}_0\ \s\s|\ \s\s
(j^+_0+j^-_0)|\phi\rangle=0\ \}
\qqq
This corresponds to taking only the highest weight vectors
belonging to the complex conjugate pairs of left-right
representations of the \s$u(1)\s$ current algebra.
${\NH}'_0\s$ is spanned by vectors \s$|p^+;p^-\rangle\s$
with \s$p^+=p^-\s$. \s
Let \s$E_0\s$ and \s$E'_0\s$ denote the orthogonal projections
of \s${\NH}\s$
on \s${\NH}_0\s$ and \s${\NH}'_0\s$, \s respectively.
Notice that
\qq
E_0\s V_{q^+q^-}(1)\s|p^+;p^-\rangle\s=
(-1)^{(q^++q^-)(p^+-p^-)/2}\s
|q^++p^+;q^-+p^-\rangle\ .
\qqq
Let us consider the (commutative) algebra \s$\CA'_0\s$
generated by operators \s$E'_0\s V_{qq}(1)\s\big|_{{\NH}'_0}\s$.
We obtain this way the effective target geometry
\qq
({\NH}'_0,(\CH+{_1\over^{12}})|_{{\NH}'_0},\CA'_0)_r
\label{eftar}
\qqq
describing the low energy regime of the sigma
model when \s$r\gg 1\s$.
\s${\NH}'_0\s$ is naturally isomorphic
to \s$L^2(S^1\hs{0.03cm},{{dx}\over^{\sqrt{2\pi}}})\s$
by identifying the vector \s$|p_0,0\rangle\equiv|p_0;p_0
\rangle\s$
with the function \s$\ee^{-ip_0x}\s$. With this identification,
\qq
({\NH}'_0,(\CH+{_1\over^{12}})
|_{\NH'_0},\CA'_0)_r\s\cong\s(\s L^2(S^1\hs{0.03cm},
dx)\s,\s -\hf\s r^{-2}{_{d^2}\over^{dx^2}}\s,
\s C^\infty(S^1)\s)\ ,
\qqq
so that the effective target (\ref{eftar}) is just
a circle of radius \s$r\s$.
Notice that we have used only the pair of \s$u(1)\s$
current algebras \s$(j^\pm_n)\s$ in order to define the effective
target geometry.
\vs 0.3cm

What is the meaning of duality in this language?
Replacing the pair \s$j^\pm\s$ of currents by
\s${\tilde j}^\pm=\pm j^\pm\s$
and repeating the whole construction
of the target geometry, we obtain a different triple
\qq
({\NH}''_0,(\CH+{_1\over^{12}})|_{{\NH}''_0},\CA''_0)_r\ ,
\qqq
where
\qq
{\NH}''_0&=&\{\ |\phi\rangle\in{\NH}_0\ \s\s|\ \s\s
(j^+_0-j^-_0)|\phi\rangle=0\ \}\ \
\qqq
and \s$\CA''_0\s$ is generated by
\s$E''_0\s V_{q(-q)}(1)\s\big|_{{\NH}''_0}\s$.
\s${\NH}''_0\s$ is spanned by vectors
\s$|0,w\rangle\s$ and is naturally isomorphic
to \s$L^2(S^1\hs{0.03cm},{{dx}\over^{\sqrt{2\pi}}})\s$
by identifying \s$|0,w\rangle\s$
with the function \s$\ee^{-iwx}\s$. Now
\qq
({\NH}''_0,(\CH+{_1\over^{12}})
|_{\NH''_0},\CA'_0)_r\s\cong\s(\s L^2(S^1\hs{0.03cm},
dx)\s,\s -\hf\s r^{2}{_{d^2}\over^{dx^2}}\s,\s C^\infty(S^1)\s)
\qqq
and the second effective target is the circle of radius
\s$r^{-1}\s$. It describes the low energy regime of
the sigma model when \s$r\ll 1\s$. As we see, duality
results in the possibility to assign to the sigma
model two different effective targets.
\vs 0.3cm

Above, we have described the procedure to define an effective
target of a sigma model based on the \s$u(1)\s$ current
algebra symmetry. On the classical level, we could
try to select the constant modes of the string by
imposing the reparametrization invariance. In the quantum theory,
we could proceed using the Virasoro
algebra. This would lead to the zero mode geometry
\s$(\NH_0,\CH|_{\NH_0},\CA_0)\s$, \s where
now \s$\NH_0\s$ is composed
of states annihilated by \s$L^\pm_n\s$ for \s$n>0\s$
and by \s$L^+_0-L^-_0\s$, \s
and \s$\CA_0\s$ is generated by the Virasoro primary fields
at \s$z_\pm=1\s$ projected to \s$\NH_0\s$. The resulting
``small space'' \s$\NH_0\s$ is bigger than \s$\NH'_0\s$
and \s$\NH''_0\s$
constructed above: \s$L^\pm_n|p^+,p^-\rangle=0\s$
for \s$n>0\s$ and the condition
\s$(L^+_0-L^-_0)|p^+,p^-\rangle=0\s$ requires
that \s$p_w\s$ or \s$w\s$ be zero. Thus, the both dual
current algebra
targets end up in the Virasoro target.
Note that the Virasoro-based construction of the zero mode
geometry may be carried out in any CFT model.
\vs 0.4cm
\vs 0.2cm
\no{\bf{2.2.\s\ \ \s$T^2\s$ target and mirror symmetry}}
\vs 0.4cm

It will be instructive to consider a slightly more involved
example of a sigma model with a flat complex one-dimensional
torus as the target. Let \s$\CT^2\equiv S^1\times S^1\s$. The complex
structure on \s$\CT^2\s$ may be defined by the complex
coordinate \s$z=x^1+Tx^2\s$, \s
where \s$x^\mu\s$ are the angle coordinates on \s$S^1\times S^1\s$
and \s$T\equiv T_1+iT_2\s$ is a complex number with
imaginary part \s$T_2>0\s$. We shall also equip
\s$\CT^2\s$ with the flat K\"{a}hler metric \s$K={R_2\over
T_2}\s dz\s d\bar z\s$, \s where \s$R_2>0\s$, \s and a closed
2-form \s${iR_1\over T_2}\s dz\wedge d\bar z=
\beta_{\mu\nu}\s dx^\mu\wedge dx^\nu\equiv
\beta\s$, \s where \s$R_1\s$
is a real number. The K\"{a}hler metric induces a Riemannian
one \s$\gamma_{\mu\nu}\s dx^\mu dx^\nu\equiv\gamma\s$.
\qq
(\gamma_{\mu\nu})\s=\s {_{R_2}\over^{T_2}}\left(\matrix{1&T_1\cr
T_1&|T|^2}\right)\ ,\hs{1.1cm}(\beta_{\mu\nu})\s=\s
\left(\matrix{0&R_1\cr -R_1&0}\right)\ .
\label{METR}
\qqq
For later convenience, we shall combine
\s$R_1\s$ and \s$R_2\s$ into the complex
number \s$R=R_1+iR_2\s$ and shall introduce matrices
\s$(d^\pm_{\mu\nu})\equiv(\gamma_{\mu\nu}\pm \beta_{\mu\nu})\s$.
\s Notice that, if \s$\omega={R_2\over
T_2}\s dz\wedge d\bar z\s$ denotes the K\"{a}hler form, then
\s$\s\int_{\CT^2}(\beta+\omega)=2R\s.\s$
As the action of the sigma model, we shall take the sum of
(\ref{action3}) and (\ref{action4}) with \s$\gamma_{\mu\nu}\s$
and \s$\beta_{\mu\nu}\s$ as above.
\vs 0.3cm

The model with the \s$\CT^2\s$ target may be treated
in full analogy to the one with the \s$S^1\s$ target.
Let us just collect the relevant formulae:
\vs 0.2cm

\no\un{The classical solutions}:
\qq
x^\mu(\sigma,\tau)&=&x^{\mu\s 0}+\gamma^{\mu\nu}(p_\nu+
\beta_{\nu\lambda}w^\lambda)\tau+w^\mu\sigma\ \cr
&+&\sum\limits_{n\not=0}{_1\over^{\sqrt{2}\s
i\s n}}\s\alpha_{n}^{\mu\s+}
\s\ee^{in\s(\sigma+\tau)}+\sum
\limits_{n\not=0}{_1\over^{\sqrt{2}\s i\s n}}\s
\alpha_{n}^{\mu\s-}\s\ee^{-in(\sigma
-\tau)}
\ .
\qqq

\no\un{The Poisson brackets}:
\qq
\{\s x^{\mu\s 0}\s,\s p_{\nu}\s\}=\delta^{\mu}_{\nu}\s,\ \ \
\{\s\alpha_{n}^{\mu\s+}\s,\s\alpha_{m}^{\nu\s+}\s\}=
in\s \gamma^{\mu\nu}\s\delta_{n+m,0}\s,\ \ \
\{\s\alpha_{n}^{\mu\s-}\s,
\s\alpha_{m}^{\nu\s-}\s\}=
in\s \gamma^{\mu\nu}\s\delta_{n+m,0}\ .
\qqq
\vs 0.2cm

\no\un{The Hilbert s}p\un{ace of states}:
\qq
{\NH}\s=\s L^2(S^1\times S^1\hs{0.03cm},
{_{dx^1dx^2}\over^{2\pi}})^\NZ\otimes\CF^+\otimes\CF^-
\qqq
with \s$p_\mu\s$ acting as \s$i{{d}\over{dx^\mu}}\s$
and the Fock space operators \s$\alpha^{\mu\s\pm}_n\s$ satisfying
\s$[\alpha^{\mu\s\pm}_n\s,
\s\alpha^{\nu\s\pm}_m]=n\s \gamma^{\mu\nu}\s\delta_{n+m,0}\s,$
$\ \s(\alpha^{\mu\s\pm}_n)^*=\alpha^{\mu\s\pm}_{-n}\s$.
\vs 0.2cm

\no\un{The chiral fields}:
\qq
X^\mu_\pm(\tau\pm\sigma)\s=\s X^{\mu\s0}_\pm+\gamma^{\mu\nu}p_\nu^\pm
\s(\tau\pm\sigma)+\sum\limits_{n\not=0}{_1\over^{in}}\s
\alpha^{\mu\s\pm}_n\s\ee^{in\s(\tau\pm\sigma)}\ ,
\qqq
where \s$X^{\mu\s 0}_\pm={1\over i}{d\over{d p^\pm_\mu}}\s$
and the left-right momenta
\qq
p^\pm_\nu\s=\s{_1\over^{\sqrt{2}}}\s(p_\nu\pm d^\pm_{\nu\lambda}
w^\lambda)\ .
\qqq
Note that the set of momenta \s\s$\{\s (p^+,p^-)\s \}\s\s$,
together with the quadratic form \s$p^+_\mu\gamma^{\mu\nu}p^+_\nu-
p^-_\mu\gamma^{\mu\nu}p^-_\nu=2p_\mu w^\mu\s$ forms an
even self-dual Lorentzian lattice.
\vs 0.2cm

\no\un{The vertex o}p\un{erators}:
\qq
V_{q^+q^-}(z_\pm)\s=\s :\ee^{\s -i\s q^+_\mu X^\mu_+
(\tau+\sigma)
\s-\s i\s q^-_\mu X_-^\mu(\tau-\sigma)}:\s c_{q^+q^-}\ ,
\qqq
where \s$q_\mu^\pm={_1\over^{\sqrt{2}}}(n_\mu\pm
d^\pm_{\mu\nu}m^\nu)
\s\s$
with \s$n_\mu,\s m^\nu\in\NZ\s$ and the cocycles
\qq
c_{q^+q^-}(p^+,p^-)=(-1)^{n_\mu w^\mu}\ .
\qqq
$V_{q^+q^+}\s$
are mutually local fields and their smeared
versions generate a non-commutative algebra \s$\CA\s$.
\vs 0.2cm

\no\un{The current and Virasoro al}g\un{ebras}:
\qq
j^\mu_\pm(\tau\pm\sigma)=\da_\sigma X^\mu_\pm(\tau\pm\sigma)
\equiv\sum\limits_{n=-\infty}^\infty j^{\mu\s\pm}_n\s
\ee^{in(\tau\pm\sigma)}\ ,\cr
L^\pm_n=\hf\sum\limits_{m=-\infty}^{\infty}\gamma_{\mu\nu}
:j^{\mu\s\pm}_m j^{\nu\s\pm}_{n-m}:\ .\hs{1.5cm}
\qqq
\vs 0.2cm

\no\un{The Hamiltonian}:
\qq
\CH\s=\s\hf\s \gamma^{\mu\nu}\s p^+_\mu p^+_\nu
\s+\s\hf\s \gamma^{\mu\nu}\s p^-_\mu p^-_\nu\s+\s
\sum\limits_{n>0}\gamma_{\mu\nu}\s\alpha^{\mu\s+}_{-n}
\s\alpha^{\nu\s+}_n\s
+\s\sum\limits_{n>0}\gamma_{\mu\nu}\s\alpha^{\mu\s-}_{-n}
\s\alpha^{\nu\s-}_n\s-{_1\over^6}\ .
\qqq
\vs 0.3cm

All that was essentially a repetition of the story
for the \s$S^1\s$ target.
As the result of the above constructions, one obtains a triple
\qq
({\NH},\CH,\CA)_{T,R}\ .
\qqq
What are the symmetries of \s$({\NH},\CH,\CA)_{T,R}\s$ or,
more generally, of the conformal sigma models parametrized
by \s$(T,R)\s$? \s
\vs 0.2cm

\un{(i)} \ The duality transformation relates the Hilbert spaces
of the \s$(T,R)\s$ and \s$(-T^{-1},-R^{-1})\s$ models
corresponding to mutually inverse matrices \s$d^\pm\s$.
It interchanges the
momenta and the winding
numbers mapping \s$|p^+;p^-\rangle\s$ to
\s$(-1)^{p_\mu w^\mu}|(d^+)^{-1}p^+;-(d^-)^{-1}p^-\rangle\s$
and it transforms \s$\alpha^{\mu\s\pm}_n\s$ to
\s$\pm{(d^\mp)^{-1}}_{\mu\nu}
\alpha^{\nu\s\pm}_n\s$. \s
It results in the equivalence
\qq
({\NH},\CH,\CA)_{T,R}\cong({\NH},\CH,\CA)_{-T^{-1},-R^{-1}}\ .
\qqq
\vs 0.2cm

\un{(ii)} \ Further symmetries are due to the fact
that complex tori
with \s$T\s$ replaced by \s$T+1\s$ or \s$-T^{-1}\s$
are conformally equivalent to the original ones (with the
same value of \s$R\s$)\s:
\qq
({\NH},\CH,\CA)_{T,R}\cong({\NH},\CH,\CA)_{T+1,R}
\cong({\NH},\CH,\CA)_{-T^{-1},R}\ .
\qqq
\vs 0.2cm

\un{(iii)} \ The change \s$R\mapsto R+1\s$ may be absorbed
in a shift of the momenta \s$p_\nu\mapsto p_\nu-
\epsilon_{\nu\lambda}w^\lambda\s$ so that
\qq
({\NH},\CH,\CA)_{T,R}\cong({\NH},\CH,\CA)_{T,R+1}\ .
\qqq
\vs 0.2cm

The above symmetries imply that
\qq
({\NH},\CH,\CA)_{T,R}\cong({\NH},\CH,\CA)_{T',R'}\ ,
\qqq
where
\qq
T'={_{aT+b}\over^{cT+d}}\s,\ \ R'={_{a'R+b'}\over^{c'R+d'}}\ ,
\qqq
for \s$\left(\matrix{a&b\cr c&d}\right),\
\left(\matrix{a'&b'\cr c'&d'}\right)\s\in\s SL(2,\NZ)\s$.
\vs 0.2cm

\un{(iv)} \ Explicit calculation shows that
\qq
\hf\s \gamma^{\mu\nu}p^+_\mu p^+_\nu\s+\s\hf\s \gamma^{\mu\nu}
p^-_\mu p^-_\nu\s=\s{_1\over^4}\s {_1\over^{T_2R_2}}\s
|Tp_1-p_2+Rw_1+TRw_2|^2\cr
+\s{_1\over^4}\s {_1\over^{T_2R_2}}\s
|Tp_1-p_2+\bar Rw_1+T\bar Rw_2|^2\ .
\qqq
It is easy to verify that the transformation relating
the spaces of states for \s$(T,R)\s$ and \s$(R,T)\s$ by
interchanging \s$p_1\s$ and \s$w_1\s$
in the vectors \s$|p^+;p^-\rangle\s$ (accompanied
by multiplication by \s$(-1)^{p_1w^1}\s$) and mapping
\s$\alpha^{1\s\pm}_n\s$ to \s\s$\pm T_2R_2^{-1}\alpha^{1\s\pm}_n
\s-\s(T_1R_2\mp T_2R_1)R_2^{-1}\alpha^{2\s\pm}_n\s\s$
establishes the equivalence
\qq
({\NH},\CH,\CA)_{T,R}\cong({\NH},\CH,\CA)_{R,T}\ .
\qqq
This is the simplest instance of the ``mirror symmetry''
\cite{GreenPless}\cite{Yau} which claims the equivalence
of conformal field
theories for complex targets with roles of
the complex and the K\"{a}hler structures interchanged.
We shall return to this topic in Sect.\s\s 5.2.
\vskip 0.3cm

Notice that the space of complex 2-dimenstional tori with
flat K\"{a}hler metric and covariantly constant closed
2-forms is \s\s$(H_+\times H_+)/SL(2,\NZ)\s$, \s where \s$H_+\s$
is the upper half-plane of \s$T\s$ and \s$R\s$ and
\s$SL(2,\NZ)\s$ acts on \s$H_+\times H_+\s$ by
\s\s$(T,R)\mapsto({aT+b\over cT+d},R)\s$. \s In contrast,
the space of the sigma models with such targets
is \s\s$(H_+\times H_+)/(SL(2,\NZ)\times SL(2,\NZ))/\NZ_2\s$,
\s where \s$\NZ_2\s$ acts by interchanging \s$T\s$ and \s$R\s$.
\vskip 0.3cm

Using the \s$u(1)\s$ currents \s$j^{\mu\s\pm}\s$ we could
again define the zero-mode restriction of the theory
by imposing the conditions \s$j^{\mu\s\pm}_n|\phi\rangle=0\s$.
Different effective toroidal target geometries with different
Riemannian metrics may then be obtained by restricting,
additionally, the zero modes. It will be convenient to introduce
the following combinations of currents:
\qq
{\tilde j}^{\mu\s\pm}_n=\pm d^{\mp}_{\mu\nu}
j^{\nu\s\pm}_n\ ,\hs{1cm}{\bf j}^{\s\pm}_n\equiv {_i\over^{T_2}}
(j^{1\s\pm}_n
+Tj^{2\s\pm}_n)\ ,\hs{1cm}
\bar{\bf j}^{\s\pm}_n\equiv -{_i\over^{T_2}}
(j^{1\s\pm}_n+\bar Tj^{2\s\pm}_n)\ .
\qqq
Then the zero mode conditions:
\vs 0.2cm

1.\ \ \hs{-0.03cm}$(j^{\mu\s+}_0+j^{\mu\s-}_0)|\phi\rangle=0\s\
$ give the
metric \s$\gamma_{\mu\nu}dx^\mu dx^\nu\s$ on the \s$\CT^2\s$ target,
\vs 0.2cm

2.\ \ $({\tilde j}^{\mu\s+}_0+{\tilde j}^{\mu\s-}_0)|\phi
\rangle=0\s\ \hs{0.23cm}$
give the dual metric \s${\tilde\gamma}_{\mu\nu}
dx^\mu dx^\nu\s$, \s where \s${\tilde\gamma}_{\mu\nu}\equiv
{(d^+)^{-1}}^{\mu\lambda}\gamma_{\lambda\rho}
{(d^-)^{-1}}^{\rho\nu}\s$,
\vs 0.2cm

3.\ \ $\hs{0.02cm}({\bf j}^{\s\pm}_0+\bar{\bf j}^{\s\mp}_0)
|\phi\rangle=0\ $ \ \ \ \ \s\hs{0.05cm}give
the mirror metric \s${\tilde{\tilde\gamma}}_{\mu\nu}dx^\mu
dx^\nu\s$ with \s$({\tilde{\tilde\gamma}}_{\mu\nu})\s$
as in (\ref{METR}), but with \hs*{5cm}$T\s$
and \s$R\s$ interchanged.
\vs 0.2cm

\no The quantum mechanics of a particle on the torus with
the above metrics describes then the low energy
spectrum of the sigma model in the regimes
\s$\gamma_{\mu\nu}\gg 1\s$, \s$\tilde\gamma_{\mu\nu}\gg 1\s$
and \s${\tilde{\tilde\gamma}}_{\mu\nu}\gg 1\s$, respectively.
Notice, that the passage from the first target to the
dual one is obtain by the interchange of \s$j^{\mu\s\pm}\s$
and \s${\tilde j}^{\mu\s\pm}\s$ and to the
mirror one by the interchange of \ ${\bf j}^-\s$
and \ $\bar{\bf j}^-\s$.
\vskip 0.4cm

The discussion of the last two sections may be
generalized to the case of general toroidal
sigma models leading to more general even self-dual
Lorentzian lattices
of left-right momenta. An interesting exercise, which remains
to be done, is the calculation of the effective
targets for the sigma models with fields taking values
in toroidal orbifolds.
\vskip 0.8cm

\nsection{\hspace{-.7cm}.\ \ WZW and coset theories}
\vs 0.2cm
\no{\bf{3.1.\s\ \ Geodesic motion on a group}}
\vs 0.4cm

The geodesic motion on \s$S^1\s$ or \s$S^1\times S^1\s$ was
essentially free and, consequently, easily solvable both
classically and in quantum mechanics. \s$S^1\s$ is the simplest
compact Lie group. For other compact Lie groups \s$G\s$,
the geodesic motion w.r.t. the invariant metric,
although not free, may be also easily
solved using its symmetries. The relevant
action functional of the particle trajectory
\s$\tau\mapsto g(\tau)\in G\s$ is
\qq
S(g(\cdot))\s=\s-{_k\over^4}\int\tr\s (g^{-1}\da_\tau g)^2
\s d\tau\ ,
\qqq
where the coupling constant \s$k>0\s$. The classical
equations
\qq
\da_\tau(g\da_\tau g^{-1})=0
\qqq
give the geodesics on \s$G\s$. The quantized system has
\s$L^2(G,dg)\s$ as the space of states, where \s$dg\s$
stands for the normalized Haar measure. It carries
two (left-right) commuting (regular) unitary
representations of \s$G\s$ acting by
\qq
L_{g_1}R_{g_2}f(g)=f(g_1^{-1}g\s g_2)\ .
\qqq
The decomposition of the regular representation into
the irreducible components results in the isomorphism
\qq
L^2(G,dg)\s\cong\s\bigoplus\limits_R V_R\otimes V_{\bar R}\ ,
\label{decomp}
\qqq
where \s$R\s$ runs through all unitary irreducible representations
of \s$G\s$ in (finite-dimensional) Hilbert spaces \s$V_R\s$
and \s$\bar R\s$ denotes the representation complex
conjugate to \s$R\s$.
\s$V_R\otimes V_{\bar R}\s$ is
spanned by matrix elements of the representation \s$\bar R\s$.
The left (right) representation of \s$G\s$
acts on the left (right) factor.
For \s$G=SU(2)\s$, we shall label
representations \s$R\s$ by spins \s$j=0,{1\over 2},1,\dots\s$.
\vs 0.2cm

Let \s$(t^A)\s$ be a basis of the Lie algebra \s${\bf g}\s$
of \s$G\s$ s.t. \s$\tr\s\s t^At^B=\hf\s\delta^{AB}\s$. Let
\s$J^{A+}={1\over i}\s dL(t^A)\s$ and \s$J^{A-}={1\over i}\s
dR(t^A)\s$
be the (selfadjoint) operators expressing the infinitesimal
actions of \s$t^A\s$ in \s$L^2(G)\s$. The quantum
Hamiltonian of the model is given by
\qq
\CH={_2\over^k}\s J^{A+}J^{A+}={_2\over^k}\s J^{A-}J^{A-}\equiv
\s-{_2\over^k}\s\Delta_G
\qqq
i.e. it is proportional to the Laplacian on \s$G\s$
which acts as multiplication by minus the quadratic Casimir
\s$c_R=c_{\bar R}\s$ in the \s$V_R\otimes V_{\bar R}\s$ subspace
of (\ref{decomp}). The triple
\qq
(\s L^2(G,dg)\s,\s-{_2\over k}\Delta_G\s,\s C^\infty(G)\s)
\label{grgeom}
\qqq
encodes the invariant Riemannian geometry on \s$G\s$.
\vs 0.4cm
\vs 0.2cm
\no{\bf{3.2.\s\ \ WZW model}}
\vs 0.4cm

What about a sigma model with the group \s$G\s$ target?
If we take
\qq
S(g(\cdot,\cdot))\s=\s-{_{k}
\over^{8\pi}}\int\tr\s\s(\s(g^{-1}
\da_\tau g)^2-(g^{-1}\da_\sigma g)^2)\s d\sigma\s d\tau
\qqq
as the action functional
then the model requires an infinite renormalization of
the coupling constant \s$k\s$ and is believed to result in
a massive two-dimensional field theory. One may,
however,
add to the action the term (\ref{action4})
with a 2-form \s$\beta\equiv\beta_{\mu\nu}\s
dx^\mu\wedge dx^\nu\s$ satisfying
\qq
d\beta={_{k}\over^3}\s\tr\s\s(g^{-1}dg)^{\wedge 3}\ .
\qqq
Such 2-forms on \s$G\s$ exist only locally. The freedom
of their choice results in the \s$2\pi k\NZ$-valued
ambiguity in the definition of the
action which is irrelevant at the classical
level but restricts the values of the coupling constant
\s$k\s$ to (positive) integers in the quantum theory.
For such values, the modified sigma model with the group
target does not require renormalization of \s$k\s$
and gives rise to the WZW
model of CFT (of ``level'' \s$k\s$) \cite{WittenBos}.
\vs 0.3cm

The Hilbert space \s${\NH}\s$ of the WZW model is built from
the representations of two
commuting copies of the Kac-Moody
algebra \s$\hat{\bf g}\s$ generated by elements \s$J_n^{A\pm}\s$
satisfying the commutation relations
\qq
[\s J^{A\pm}_n\s,\s J^{B\pm}_m\s]\s=\s
i\s f^{ABC}\s J^{C\pm}_{n+m}\s+\s\hf\s kn\s\delta^{AB}\s
\delta_{n+m,0}\ ,
\qqq
where \s$f^{ABC}\s$ are the structure constants:
\s$[t^A,\s t^B]=i\s f^{ABC}t^C\s$. \s
$\hat{\bf g}\s$ is the central extension of the loop algebra
\s$L{\bf g}\s$ and \s$k\s$ is its central charge (level). \s
The unitary (\s$(J^A_n)^*=J^A_{-n}\s$) irreducible highest weight
representations of \s$\hat{\bf g}\s$
are labeled by \s$k\s$ (a non-negative integer)
and an irreducible representation \s$R\s$ of \s$G\s$
from a restricted class (for \s$G=SU(2)\s$, \s with spin
\s$j\leq{k\over 2}\s$) \s whose elements we shall call
integrable (at level \s$k\s$) \cite{KacV}.
They act in infinite-dimensional
(if \s$k>0\s$) \s spaces \s${\bf V}^k_R\s$.
For \s$G\s$ connected and simply connected
\qq
{\NH}\s\s=\s\s\{\bigoplus\limits_{R\atop{\rm integrable}}
{\bf V}^k_R\otimes {\bf V}^k_{\bar R}\ \s\}^-\ ,
\qqq
where \s$\{\cdots\}^-\s$ denotes the Hilbert space completion.
The current modes \s$J^{A\pm}_n\s$'s generate,
by the Sugawara construction,
two commuting representations of the Virasoro algebra
\qq
L^\pm_n\s=\s{_1\over {k+g^\vee}}\sum\limits_{m=-\infty}^{\infty}
:J^{A\pm}_m\s J^{A\pm}_{n-m}:
\qqq
with central charge \s$c^G_k={k\s{\rm dim}(G)\over k+g^\vee}\s$
\s(\s$g^\vee\s$ denotes the quadratic Casimir of the adjoint
representation). The Hamiltonian of the model is
\qq
\CH=L^+_0+L^-_0-{_1\over^{12}}c^G_k\ ,
\qqq
whereas
\qq
\CP=L^+_0-L^-_0
\qqq
generates the space translations.
\vs 0.3cm

Let us define a small space of states \s${\NH}_0\s$
as composed of vectors \s$|\phi\rangle\in\NH\s$
satisfying the highest weight condition \s$J^{A\pm}_n\s
|\phi\rangle=0\s$ for \s$n>0\s$. \s${\NH}_0\s$
carries the representation of \s${\bf g}\oplus{\bf g}\s$
(and of \s$G\times G\s$) \s given by the action of
\s$J^{A\pm}_0\s$'s\s. \s
With respect to this representation,
\qq
{\NH}_0\s\s\cong\bigoplus\limits_{R\atop{\rm integrable}}
V_R\otimes V_{\bar R}
\label{DC}
\qqq
which may be naturally identified (see (\ref{decomp})\s)
\s with a subspace of
\s$L^2(G)\s$ which we shall denote \s$L^2_k(G)\s$.
\s The Hamiltonian \s$\CH\s$ preserves \s${\NH}_0\s$
and reduces there to \s$-{2\over {k+g^\vee}}\Delta_G-{1\over 12}
c^G_k\s$.
To each vector \s$|\phi\rangle\in{\NH}_0\s$ there
corresponds a unique primary field
\s$V_{|\phi\rangle}(z_\pm)\s$ from a local family such that
\qq
[\s J^{A\pm}_n\s,\s V_{|\phi\rangle}(z_\pm)\s\s]\s\s=\s\s
z_\pm^n\s\s V_{J^{A\pm}_0|\phi\rangle}(z_\pm)\s\s
\qqq
and
\qq
\lim_{z_\pm\rightarrow 0}\s V_{|\phi\rangle}
(z_\pm)\s\s|{\rm vac}\rangle\s=\s|\phi\rangle\ .
\qqq
$V_{|\phi\rangle}(z_\pm)\s$ is also a primary field
of the Sugawara Virasoro algebras (see (\ref{Virpf})).
Taking the algebra \s$\CA\s$ generated by the
primary fields, we obtain a triple \s$({\NH},\CH,\CA)\s$
encoding the geometry of the WZW CFT.
\vs 0.3cm

We have presented the WZW theory as a sigma model with
a group target, but its construction on the quantum level
has proceeded directly through the representation theory
of the Kac-Moody algebra. Again the question arises how
to see the target manifold geometry in the resulting
quantum theory. It seems reasonable to proceed via the
restriction to the ``zero mode'' subspace \s${\NH}_0\s$
of primary states with the Hamiltonian
\s$\CH_0=(\CH+{1\over12}c^G_k)|_{{\NH}_0}\s$.
Such a restriction cuts out the descendent states created by
operators \s$J^{A\s\pm}_n\s$, \s with \s$n<0\s$, \s which
increase the energy by \s$|n|\s$. Since, for \s$G\s$ simple
and simply connected, the left-right
representations of the current algebra \s$\hat{\bf g}\s$ appear
only in complex conjugate pairs (unlike for the
\s$S^1\s$ sigma model), we shall not impose any further zero
mode conditions on the small space of states. In fact,
\s$\NH_0\s$ contains
all states with low energy (\s$\ll 1\s$) for \s$k\gg 1\s$.
\s To encode the effective target geometry, we still need an
algebra \s$\CA_0\s$ of ``functions'' on the target.
Let \s$E_0\s$ denote the orthogonal projection
of \s${\NH}\s$ onto \s${\NH}_0\s$. We shall
take as \s$\CA_0\s$ the algebra generated by operators
\s$E_0 V_{|\phi\rangle}(1)\s\big|_{{\NH}_0}\s$.
This way we obtain a triple
\qq
({\NH}_0,\CH_0,\CA_0)\ .
\qqq
What is its relation to the triple (\ref{grgeom})
representing the Riemannian geometry of the group \s$G\s$?
As we have already noticed, \s${\NH}_0\s$ is a subspace
of \s$L^2(G)\s$ and \s$\CH_0\s$ coincides on it with
\s$-{2\over {k+g^\vee}}\Delta_G\s$. Thus we only need to understand
the relation of \s$\CA_0\s$ to the algebra of functions
on \s$G\s$.
\vs 0.3cm

Let \s$f_l\s,\ l=1,2,3\s,$ \s be
three functions on \s$G\s$ lying, in the decomposition
(\ref{decomp}),
inside \s$V_{R_l}\otimes V_{\bar R_l}\s$, \s respectively, where
\s$R_l\s$ are representations integrable at level \s$k\s$.
\s$f_l\s$ define vectors \s$|f_l\rangle\in\NH_0\s$.
\s Let \s\s${\rm Inv}_{(R_l)}\s\s$ denote
the subspace of \s\s$\otimes_lV_{R_l}\s\s$ invariant with
respect to the diagonal action of \s$G\s$.
The matrix elements of the primary
fields \s$V_{|\phi\rangle}(1)\s$ between
the states in \s${\NH}_0\s$ have the following form
\qq
\langle \overline{f_{3}}|\s V_{|f_{2}\rangle}(1)\s|f_{1}\rangle\s
=\s \langle C^k_{(R_l)}\s|\otimes_l\hs{-0.05cm} f_l\rangle\ ,
\label{matel}
\qqq
where \s\s$C^k_{(R_l)}
\in{\rm Inv}_{(R_l)}
\otimes{\rm Inv}_{(\bar R_l)}\cong{\rm End}({\rm Inv}_{(R_l)})\s$
is a positive element. \s\s For \s \s$G=SU(2)\s$,
\s\s${\rm Inv}_{(j_l)}
\otimes{\rm Inv}_{(\bar j_l)}\s\s$, \s if non-vanishing,
is canonically isomorphic
to \s$\NC\s$ and formula (\ref{matel}) takes the simpler
form
\qq
\langle \overline{f_{3}}|\s V_{|f_{2}\rangle}(1)\s|f_{1}\rangle\s
=\s C^k_{(j_1,j_2,j_3)}\s\int\limits_G\prod\limits_l{f_l(g)}
\s\s dg\ ,
\label{matel1}
\qqq
where \s$C^k_{(j_l)}\s$ are (up to normalization) the
operator product coefficients \cite{BPZ}
for the WZW theory. They have been computed in \cite{ZamFat}:
\qq
C^k_{(j_1,j_2,j_3)}\s=\s(J+1)!\ P(J+1)\s\s
P(1)^{1/2}\s\s\prod\limits_{l=1}^3{P(\hat{j}_l)\s\s\hat{j}_l!
\over(2j_l+1)^{1/2}\s(2j_l)!\s\s P(2j_l)^{1/2}\s P(2j_l+1)^{1/2}}\ ,
\label{opprod}
\qqq
where \s\s$J\equiv j_1+j_2+j_3\s,$\ \ $\hat{j}_l
\equiv J-2j_l\s\s$ and
\qq
P(0)\s=\s 1\s,\ \ \ \ P(j)\s=\s\prod\limits_{i=1}^j\s{\Gamma
({i\over k+2})\over\Gamma(1-{i\over k+2})}\ .
\qqq
In (\ref{opprod}), it is assumed that \s$j_l\leq{k\over 2}\s$.
\vs 0.2cm

As we see from the relation (\ref{matel1}), for \s$G=SU(2)\s$,
\s$\CA_0\s$ is a deformed version of the algebra of multiplication
by functions from \s${\NH}_0\s$. \s It is, however,
non-commutative (in general). The commutativity is restored
in the classical limit \s$k\rightarrow\infty\s$, \s where
\s${\NH}_0\rightarrow L^2(G)\s$ and \s$C^k_{(j_1,j_2,j_3)}
\rightarrow 1\s$. \s Also for general \s$G\s$, \s
\qq
\lim_{k\rightarrow\infty}C^k_{(R_l)}\s=\s{\rm I}\s\s\in\s
{\rm End}({\rm Inv}_{(R_l)})
\label{LIM}
\qqq
so that
$$\lim_{k\rightarrow\infty}\langle C^k_{(R_l)}\s
|\otimes_l\hs{-0.05cm} f_l\rangle\s=\s
\int\limits_G\prod\limits_l{f_l(g)}\s\s dg$$
and one recovers, in the \s$k\rightarrow\infty\s$ limit,
the triple (\ref{grgeom}) encoding the commutative geometry
of the group manifold. For finite \s$k\s$, \s$({\NH}_0,
\CH_0,\CA_0)\s$ represents a finite (\nobreak\s\nobreak$
{\NH}_0\s$
is finite-dimensional) non-commutative geometry of the
effective target of the WZW model.
\vskip 0.4cm

We should stress once more that the infinite
symmetry of the conformal model is an important
input in our definition of the effective target.
If, for example, in the case of the WZW model with \s$G=SU(2)\s$
we have used only the \s$u(1)\times u(1)\s$ current algebra,
we should have obtained a different effective target.
In particular, the \s$u(1)\s$ target would
coincide for the level \s$k=1\s$
with that of the sigma model with field values in \s$S^1\s$
of radius \s$r=1\s$.

\vs 0.4cm
\vs 0.2cm
\no{\bf{3.3.\s\ \ Coset quantum mechanics}}
\vs 0.4cm

There is a simple way, called the coset construction
\cite{GKO}, to generate new CFT's
from the group $\s G$-valued WZW model.
Let us start by describing the quantum mechanical counterpart
of the coset construction. As we have seen, \s$L^2(G)\s$
carries a unitary representation of \s$G\times G\s$.
For any subgroup \s$H\subset G\times G\s$, we may
consider the subspace \s$L^2(G)_H\subset L^2(G)\s$ of functions
invariant under \s$H\s$. If \s$H\s$ is a subgroup
of \s$G_{\rm left}\s$ or a subgroup of \s$G_{\rm right}\s$
or a product of such subgroups, we end up with the
space of square-integrable functions on the (left, right
or left-right) coset space. There is another possibility
which has attracted less attention by harmonic analysts. One
may take \s\s$H\subset G_{\rm diag}\subset G
\times G\s$. \s\s This leads to functions
on \s$G/{\rm Ad}(H)\s$. \s Let us decompose the irreducible
representations of \s$G\s$ with respect to \s$H\s$:
\qq
V_{{R}}\s\cong\s\bigoplus\limits_{{r}}
M^{{R}}_{{r}}\otimes V_{{r}}\ ,
\qqq
where \s$R\s$ refers to irreducible representations
of \s$G\s$ and \s$r\s$ to those of \s$H\s$. The multiplicity
spaces are \s$M^{R}_r={\rm Hom}_H(V_r,V_R)\s$. Since
the space \s${\rm Inv}_{r,r'}\s$
of vectors in \s$V_r\otimes V_{r'}\s$
invariant under the diagonal action of \s$H\s$
is canonically isomorphic to \s$\delta_{\bar r,r'}\s\NC\s$,
\s the decomposition (\ref{decomp}) reduces to
\qq
L^2(G)_H\s\cong\s\bigoplus\limits_{{R},{r}}
M^{R}_{r}\otimes
{\overline{M}}^{R}_{r}\ .
\label{decomp1}
\qqq
Let \s${{{\bf h}^\perp}}\s$ be the orthogonal
complement of \s${\bf h}\s$
in \s$\bf g\s$. We shall choose the basis \s$(t^A)=((t^{\rm a}),
(t^{\alpha}))\s$ of \s$\bf g\s$ so that \s$t^{\rm a}\s$
(\s$t^{\alpha}\s$) \s span \s$\bf h\s$ (\s${{\bf h}^\perp}\s$)\s.
Both \s$-\Delta_G\equiv J^{A\pm}J^{A\pm}\s$ and
\s$-\Delta^\pm_H\equiv J^{{\rm a}\pm}J^{{\rm a}\pm}\s$
commute with the generators \s$J^{{\rm b}\pm}\s$
of the left and right \s${\bf h}\s$ symmetry, so that we may take
\s$-{2\over k}(\Delta_G-\Delta^\pm_H)={2\over k}
J^{\alpha\pm}J^{\alpha\pm}\s$
as the Hamiltonian of the reduced system (both
signs give the same operator on \s$L^2(G)_H\s$)\s. \s Finally,
we may take the algebra \s$C^\infty(G)_H\s$ of multiplication
by the functions invariant under the adjoint action of \s$H\s$
and consider the triple
\qq
(\s L^2(G)_H\s,\s-{_2\over^k}(\Delta_G-\Delta^\pm_H)\s,\s
C^\infty(G)_H\s)
\label{cosettri}
\qqq
representing what we shall, somewhat abusively, call the
coset geometry.
\vs 0.3cm

Is the coset geometry a standard
Riemannian one? This is not so and there are
two reasons for it. First, since the adjoint
action is not free, \s$G/{\rm Ad}(H)\s$ might not
be a smooth manifold. In particular,
for \s$H=G\s$, \s$L^2(G)_G\s$ is the space of class
functions on \s$G\s$ spanned by the characters
of irreducible representations. \s$G/{\rm Ad}(G)\s$ may
be identified with the orbifold \s$T/W\s$, \s where \s$T\s$
is the Cartan subgroup of \s$G\s$ and \s$W\s$ is the Weyl
group. Second, \s$-{2\over k} (\Delta_G-\Delta_H^\pm)\s$
differs from (\nobreak\s\nobreak$-\hf$ times\s) the
Laplace-Beltrami operator
for the (in general
singular) metric \s$\gamma\s$ on \s$G/{\rm Ad}(H)\s$
which may be extracted from the triple (\ref{cosettri})
by the procedure
described in Introduction. The two operators are, essentially,
different quantizations of the same classical energy of
a geodesic motion on \s$G/{\rm Ad}(H)\s$.
Let us explain the last point in more details.
A similar situation in the context of gauge theories
has been analyzed in \cite{Gaw0}.
\vs 0.3cm

The quantum mechanical system (\ref{cosettri}) may be
obtained by quantization of a classical one
which couples the geodesic motion of the particle
on the group \s$G\s$ to \s${\bf h}$-valued
gauge fields \s$\tau\mapsto A_\pm(\tau)\s$. The action
functional for the coupled system is
\qq
S(g(\cdot),A_\pm(\cdot))&=&-{_k\over^4}\int\tr\s\s(g^{-1}
\da_\tau g)^2\s d\tau\cr
&&\s+\s{_k\over^2}\int\tr\s\s(\s(g\s\da_\tau g^{-1})
A_-+A_+ (g^{-1}\da_\tau g)+gA_+
g^{-1}A_--A_+A_-\s)\s\s d\tau\ \s
\qqq
and is invariant under the gauge transformations
\s$g\mapsto hgh^{-1}\s,$ \s$A_\pm\mapsto hA_\pm h^{-1}
\hs{-0.04cm}+h\da_\tau h^{-1}\s$ for arbitrary
\s$H$-valued \s$h(\tau)\s$.
Note that the gauge field enters quadratically and may
be easily eliminated in the functional integral leading
to the effective action \cite{GawKup2}
\qq
S_{\rm eff}(g(\cdot))\s=\s-{_k\over^4}\int
\tr\hs{-0.2cm}&[&\hs{-0.25cm}
(g^{-1}\da_\tau g)\s(g^{-1}\da_\tau g)\ \s\cr
&+&2(g^{-1}\da_\tau g)
\s\s(1-E\s{\rm Ad}_g\s E)^{-1}E\s{\rm Ad}_g\s
(g^{-1}\da_\tau g)\s]\s\s d\tau\ ,
\label{MET}
\qqq
where \s$E\s$ is the orthogonal projection
in \s${\bf g}\s$ onto \s${\bf h}\s$ and \s$E^\perp=1-E\s$.
The effective action is invariant under
the transformations \s$g\mapsto h
gh^{-1}\s$, \s again with \s$h(\tau)\s$ taking values in
\s$H\s$, \s so that it defines the geodesic flow on
\s$G/{\rm Ad}(H)\s$ with respect to a (in general singular)
metric which may be easily read off (\ref{MET}).
Viewing \s${1\over i}g^{-1}\da_\tau g\equiv X\s$
modulo \s${\rm Ad}_{g^{-1}}(Y)-Y\s$ with \s$Y\in{\bf h}\s$
as representing vectors tangent to \s$G/{\rm Ad}(H)\s$ at point
\s$[g]\s$, we obtain for their length squared
\qq
\|X\|^2\s=\s{_k\over^2}\s\tr\s\s[\s
X^2\s+\s 2X\s(1-E\s{\rm Ad}_g\s E)^{-1}E\s{\rm Ad}_g\s
X\s]\ .
\label{MTR}
\qqq
The dual metric on the cotangent bundle of \s$G/{\rm Ad}(H)\s$
is given by a simpler expression. If \s$X'\in{\bf g}\s$
\s s.t. \s$E({\rm Ad}_g(X')-X')=0\s$
represents a covector \s${1\over i}\s\tr\s X'g^{-1}dg\s$
tangent to \s$G/{\rm Ad}(H)\s$ at
\s$[g]\s$ then
\qq
\|X'\|^2\s={_2\over^k}\s\tr\s (E^\perp X')^2=
{_2\over^k}\s(\s\tr\s{X'}^2-\tr\s X'EX')\ ,
\label{MTRD}
\qqq
which is the classical version of the quantum
expression \s$-{2\over k}(\Delta_G-\Delta_H^-)\s$
for the coset Hamiltonian. In fact, classically,
the coset mechanics is the Hamiltonian reduction \cite{Sternberg}
of the particle on \s$G\s$ with Hamiltonian
(\ref{MTRD}) by the adjoint action of \s$H\s$.
Notice that vanishing of \s$E({\rm Ad}_g(X')-X')\s$
implies that
\qq
EX'\ \ &=&(1-E\s{\rm Ad}_g\s E)^{-1}E\s{\rm Ad}_g\s E^\perp X'
\qqq
so that we may parametrize
the covectors tangent to
\s$G/{\rm Ad}(H)\s$ by elements \s$E^\perp X'\in{\bf h}^\perp\s$.
In particular, the (imaginary)
covectors \s$\psi^{\alpha-}\s$ corresponding to
\s$E^\perp X'=2i{\sqrt{k\over 2}}t^\alpha\s$ span
the space cotangent vectors to \s$G/{\rm Ad}(H)\s$ at point
$\s[g]\s$. Explicitly,
\qq
\psi^-\equiv t^\alpha\psi^{\alpha-}=\s
{\sqrt{_k\over^2}}\s\s E^\perp\hs{-0.04cm}
\left(g^{-1}dg\s+\s{\rm Ad}_{g^{-1}}
(1-E\s{\rm Ad}_{g^{-1}}\s E)^{-1}E(g^{-1}dg)\right)\ .
\label{bASEL}
\qqq
Changing \s$g\mapsto g^{-1}\s$ we obtain another basis
\qq
\psi^+\equiv t^\alpha\psi^{\alpha+}=\s
{\sqrt{_k\over^2}}\s\s E^\perp\hs{-0.04cm}
\left((dg)g^{-1}\s+\s{\rm Ad}_{g}
\s(1-E\s{\rm Ad}_{g}\s E)^{-1}E((dg)g^{-1})\right)\ .
\label{bASER}
\qqq
The \s${\bf h}^\perp$-valued 1-forms (\ref{bASEL})
and (\ref{bASER}) transform covariantly under the adjoint
action of \s$H\s$ on \s$G\s$ and vanish on vectors tangent
to the orbits of that action. As we shall see below
in Sect.\s\s 4.5, they appear naturally in the supersymmetric
version of the coset mechanics.
Due to the simple form (\ref{MTRD}) of the dual
metric, also the tangent spaces to \s$G/{\rm Ad}(H)\s$
may be identified with \s${\bf h}^\perp\s$.
\vs 0.3cm

Let \s$d[g]\s$ denote the
volume element on \s$G/{\rm Ad}(H)\s$ determined
by the metric (\ref{MTR}). There exists a (possibly singular)
function \s$\sigma\s$ on \s$G/{\rm Ad}(H)\s$
such that for \s$f\in L^2(G)_H\s$,
\qq
\int\limits_G|f|^2dg\s=\s\int\limits_{G/{\rm Ad}(H)}
|f|^2\s\ee^{\sigma}\s d[g]
\qqq
(\s$\ee^\sigma\s$ measures, in a sense, a volume of
orbits of the adjoint action of \s$H\s$)\s.
Now
\qq
\langle f|-{_2\over^k}(\Delta_G-\Delta_H)|f\rangle
&=&{_2\over^k}\int\limits_G(|J^Af|^2-|J^{\rm a}f|^2)\s dg\cr
&=&\hf\int_{G/{\rm Ad}(H)}\|df\|^2\s\ee^{\sigma}\s
d[g]\ ,
\qqq
where \s$\|df([g])\|^2\s$ is given by eq.\s(\ref{MTRD}).
As we see, \s$-{2\over k}(\Delta_G-\Delta_H)\s$
differs from (\s$-{1\over 2}$ times) the Laplace-Beltrami
operator on \s$G/{\rm Ad}(H)\s$ for metric (\ref{MTR})
by the replacement of the Riemannian volume
\s$dv_\gamma\s$ by \s$\ee^\sigma dv_\gamma\s$, \s both in the
definition of the \s$L^2\s$ scalar product and in the Dirichlet
form. This has been noticed in \cite{WittenBH}, see also
\cite{DijVV}, for a specific non-compact coset. The function
\s$\sigma\s$ is often called a ``dilaton'' in the physical
literature. It is a quantum-mechanical effect: it adds
a correcting potential of order \s$1\over k^2\s$ to the
Laplace-Beltrami operator \cite{Gaw0}.
\vs 0.4cm
\vs 0.2cm
\no{\bf{3.3.\s\ \ Coset CFT models}}
\vs 0.4cm

The coset  construction is the
field theory version of the quantum-mechanical reduction
by \s$H\subset G_{\rm{diag}}\subset G\times G\s$.
The original idea of \cite{GKO} was to decompose a
representation of the affine Kac-Moody algebra
\s$\hat{{\bf g}}\s$ with respect
to a subalgebra
\s$\hat{{\bf h}}\s$
and to look at the multiplicity spaces.
If \s${\bf V}^k_{R}\s$ carries the irreducible unitary
representation of \s$\hat{\bf g}\s$ then
\qq
{\bf V}^k_{R}\ \cong\bigoplus\limits_{r\atop
{\rm integrable}}{\bf M}^{{k,R}}_{r}\s\otimes\s{\bf V}
^{k}_{r}\ ,\label{multiplic}
\qqq
where \s${\bf V}^{k}_{r}\s$ are the spaces of
irreducible unitary representations of \s$\hat{{\bf h}}\s$.
\s The key point is that the
multiplicity spaces
\s${\bf M}^{{k,R}}_{r}\s$ carry unitary
representations of the Virasoro algebra obtained
by taking the difference of the two Sugawara constructions,
relative to \s$\hat{{\bf g}}\s$ and relative
to \s$\hat{{\bf h}}\s$:
\qq
L^{{\rm cs}\s\pm}_n\s
=\s{_1\over^{k+g^\vee}}\s\sum\limits_m:J_{m}^{A\pm}\s
J_{n-m}^{A\pm}:
\s-\s{_1\over^{k+h^\vee}}\s\sum\limits_m:J_{m}^{{\rm a}\pm}
\s J_{n-m}^{{\rm a}\pm}:\ .
\qqq
The operators \s$L^{{\rm cs}\s\pm}_n\s$ commute with the currents
\s$J_{n}^{{\rm b}\pm}\s$ so that,
indeed, they act in \s${\bf M}^{{k,R}}_{r}\s$. \s They
form a representation of the Virasoro algebra with central
charge \s$c^{G/H}\equiv c^G_k-c^H_{k}\s$, \s the difference of
the two Sugawara central charges.
\vskip 0.3cm

The Hilbert space of the coset theory is the (norm completed)
left-right combination of the multiplicity spaces
\s${\bf M}^{{k,R}}_{r}\s$:
\qq
\NH\s\s=\{\bigoplus\limits_{R,r\atop{\rm integrable}}
{\bf M}^{{k,R}}_{r}\otimes
{\bf M}^{{k,\bar R}}_{\bar r}\s\s\}^-\ .
\qqq
Equivalently, we may set
\qq
\NH\s=\s\{\s\ |\phi\rangle\in\NH_{_{\rm WZW}}\
\ |\ \ J^{{\rm a}\pm}_{n}\s|\phi
\rangle=0\ \s{\rm for}\s\ n>0\s,\ \s(J^{{\rm a}+}_{0}
+J^{{\rm a}-}_{0})
|\phi\rangle=0\ \s\}\ ,
\label{subst}
\qqq
where \s$\NH_{_{\rm WZW}}\s$ denotes the space of states of the
group \s$G\s$ level \s$k\s$ WZW model.
\s$\CH=L^{{\rm cs}\s+}_0+L^{{\rm cs}\s-}_0-{1\over 12}c^{G/H}\s$
defines the Hamiltonian on \s$\NH\s$.
\vs 0.2cm

To each element \s$|\phi\rangle\in \NH\s$ annihilated
by \s$L^{{\rm cs}\s\pm}_n\s$ for \s$n>0\s$ there
corresponds a primary field (out of a local family)
\s$V^{\rm cs}_{|\phi\rangle}(z_\pm)\s$ of the Virasoro
algebras \s$(L^{{\rm cs}\s\pm}_n)\s$ with the property
that
\qq
\lim_{z_\pm\rightarrow 0}V^{\rm cs}_{|\phi\rangle}(z_\pm)\s
|{\rm vac}\rangle\s=\s|\phi\rangle\ .
\qqq
Taking the algebra \s$\CA\s$ generated by
these primary fields, we obtain the triple
\s$(\NH,\CH,\CA)\s$ characterizing the geometry of the
coset CFT.
\vs 0.3cm

On the Lagrangian level, the coset construction corresponds
to gauging of a subgroup \s$H\s$ of the global
\s$G\times G\s$ symmetry in the WZW theory
\cite{BarRabSar}\cite{GawKup1}\cite{GawKup2}\cite{KarParSchYa}.
Most of the subgroups would lead
to gauge anomalies, but \s$H\subset G_{\rm diag}\s$
do not. In the diagonal case, the gauged WZW action is given
by the formula
\qq
S(g(\cdot),A_\pm(\cdot))&=&S_{\rm WZW}(g)\cr
+&&\hs{-0.5cm}{_k\over^{4\pi}}\int
\tr\s\s(\s(g\s\da_+g^{-1})A_-+A_+(g^{-1}\da_-g)+gA_+g^{-1}A_-
-A_+A_-\s)\s\s d\sigma\s d\tau\ ,
\label{gauact}
\qqq
where \s$\da_\pm=\da_\tau\pm\da_\sigma\s$ and
\s$A_\pm(\sigma,\tau)\s$ are the components of an
\s${\bf h}$-valued gauge field which, again,
enters quadratically and may be eliminated from
the functional integral.
\vs 0.4cm

The above presentation of the coset models suggests
a natural way to associate to it an effective target
geometry. As the small space of states we may take
$$\NH_0\s=\s\{\s\ |\phi\rangle\in\NH_{_{\rm WZW}}\
\ |\ \ J^{A\pm}_{n}\s|\phi
\rangle=0\ \s{\rm for}\s\ n>0\s,\ \s(J^{{\rm a}+}_{0}
+J^{{\rm a}-}_{0})
|\phi\rangle=0\ \s\}$$
which is defined using the current algebra. Notice
that \s$\NH_0\cong L^2_{k}(G)_H\s$ which
is the subspace of \s$L^2(G)_H\s$ obtained by
summing in (\ref{decomp1}) only over the
representations integrable at level \s$k\s$.
\s The restricted Hamiltonian
$$\CH_0=(\CH+{_1\over^{12}}c^{G/H})
|_{\NH_0}=-{_2\over^{k+g^\vee}}
\Delta_G+{_2\over^{k+h^\vee}}\Delta^\pm_H\ .$$
Let \s$E_0\s$ denote, as usual, the orthogonal projection
onto \s$\NH_0\s$. \s
We may generate the small algebra \s$\CA_0\s$ by operators
\s$E_0\s V^{\rm cs}_{|\phi\rangle}(1)\s|_{\NH_0}\s$
for \s$|\phi\rangle\in\NH_0\s$.
\s Their action may be explicitly described in the following
way. Let \s$f_l\s,\ \s l=1,2,3\s,\s$ be three functions
in \s$M^{R_i}_{r_i}\otimes M^{R_i}_{r_i}\subset
L^2_{k}(G)_H\s$, \s see (\ref{decomp1}).
Recall that \s$M^R_r={\rm Hom}_H(V_r,V_R)
\subset{\rm Hom}(V_r,V_R)\s$.
\s Let \s\s$C^k_{(R_l)}\s$
be the vector of the operator product
coefficients in \s\s${\rm Inv}_{(R_l)}\otimes
{\rm Inv}_{(\bar R_l)}\s\s$ giving the matrix elements
of the group \s$G\s$ primary fields, see (\ref{matel}),
and \s\s$c^k_{(r_l)}\in{\rm Inv}_{(r_l)}
\otimes{\rm Inv}_{(\bar r_l)}\s$ the one
for the group \s$H\s$. Let \s\s$\tilde c^k_{(r_l)}\in
{\rm Inv}_{(r_l)}
\otimes{\rm Inv}_{(\bar r_l)}\s\s$ be the element obtained
by inverting \s$c^k_{(r_l)}\s$ viewed as an element of
\s${\rm End}({\rm Inv}_{(r_l)})\s$ on its image.
\s$\tilde c^k_{(r_l)}\s$ is set to zero on the kernel
of \s$c^k_{(r_l)}\s$ which is orthogonal to its image.
Matrix elements of the fields
\s$V^{\rm cs}_{|\phi\rangle}(1)\s$
between states in \s$\NH_0\s$ are given by
\qq
\langle \overline{f_3}|\s V^{\rm cs}_{|f_2\rangle}(1)\s|f_1\rangle
\s=\s \langle C^k_{(R_l)}\s|
(\otimes_l\hs{-0.05cm}f_l)(\tilde c^k_{(r_l)})\rangle\ ,
\label{matel2}
\qqq
where \s$\otimes_l\hs{-0.05cm}f_l\s$ is viewed as an element
of \s\s${\rm Hom}((\otimes_lV_{r_l})\otimes(\otimes_lV_{\bar r_l})\s,
\s(\otimes_lV_{R_l})\otimes(\otimes_lV_{\bar R_l}))\s$.
Eq.\s\s(\ref{LIM}) implies then that
\qq
\lim_{k\rightarrow\infty}\s
\langle \overline{f_3}|\s V^{\rm cs}_{|f_2\rangle}(1)\s|f_1\rangle
\s=\s \int_G\prod_lf_l(g)\ dg\ .
\qqq
Hence \s$(\NH_0,\CH_0,\CA_0)\s$ encodes a finite
non-commutative geometry
deforming the coset geometry given by the triple
(\ref{cosettri}). Again the deformation disappears in the
classical limit \s$k\rightarrow\infty\s$.
\vs 0.2cm

In the special case when \s$H=G\s$, \s$L^2(G)_G\s$
is the space of class functions on \s$G\s$. \s We may
take \s$f_l\s$ to be the characters \s$\chi_l\s$ of
integrable representations \s$R_l\s$. It is easy to see
then from (\ref{matel2}) that
\qq
\langle \overline{\chi_3}|\s V^{\rm cs}_{|\chi_2\rangle}(1)
\s|\chi_1\rangle\s=\s{\rm rank}(C^k_{(R_l)})\ ,
\qqq
where \s$C^k_{(R_l)}\s$ is viewed as an element
of \s${\rm End}({\rm Inv}_{(R_l)})\s$. \s
These are the so called Verlinde dimensions \cite{Verl}
which in the limit \s$k\rightarrow\infty\s$ tend to the
dimensions of \s${\rm Inv}_{(R_l)}\s$.
\vs 0.3cm

One may envisage an alternative procedure, based
on the coset Virasoro algebra rather than on
the current algebra, for extracting
the effective target geometry from the coset theory.
It would be based on the following small Hilbert space :
\qq
{\NH}'_0\s=\s\{\ \ |\phi\in\NH\ \ |\ \ L^{{\rm cs}\s\pm}_{n}\s|\phi
\rangle=0\ \s{\rm for}\ \s n>0\s,\ (L^{{\rm cs}\s+}_0
-L^{{\rm cs}\s-}_0)\s|\phi\rangle=0\ \ \}
\qqq
with the small Hamiltonian \s$\CH'_0=(\CH+{1\over12}c^{G/H})
|_{{\NH}'_0}\s$. \s The small algebra \s$\CA'_0\s$
will then\s  be\s  generated by operators \s$E'_0\s
V^{\rm cs}_{|\phi\rangle}(1)|_{{\NH}'_0}\s$ with
\s$|\phi\rangle\in{\NH}'_0\s$. Notice that
\s$\NH_0\subset\NH'_0\s$.
\vs 0.3cm

The classical example of a coset theory \cite{GKO} is obtained
by choosing $\s G=SU(2)\times SU(2)\s$ with $\s H\s$ equal to the
diagonal $\s SU(2)\s$. Since $\s G\s$ is not simple,
the coupling constants
for the WZW theory may be chosen independently for both
$\s SU(2)\s$ factors and one sets them to $\s k\s$ and to $\s 1\s$,
\s respectively. The
coset models obtained this way give the unitary theories in the
series of {\bf minimal models} of
\cite{BPZ}. The Hilbert space of
these theories is the diagonal
combination of the irreducible unitary representations
\cite{FQS} of the Virasoro algebras \s$(L^{{\rm cs}\s\pm}_n)\s$
with central charge \s$c^{\rm min}_{k+2}=1-{6\over^{(k+2)(k+3)}}\s$.
\s\s$SU(2)\times SU(2)\ni(g_1,g_2)\mapsto\s\tr\s g_1\s\s$
defines a simple function \s$f\s$ on \s$SU(2)\times SU(2)\s$
invariant under the adjoint  action of the diagonal \s$SU(2)\s$
subgroup. The corresponding primary coset field
\s$V^{\rm cs}_{|f\rangle}(z_\pm)\s$, \s labeled
\s$\phi_{22}(z_\pm)\s$ in \cite{BPZ}, has conformal
weights \s$\Delta^\pm_{22}={3\over 4(k+2)(k+3)}\s$.
An interesting question is what is the limit of
the effective target geometries
\s$(\NH'_0\s,\s\CH'_0\s,\s\CA'_0\s)\s$ for the minimal models
when \s$k\rightarrow\infty\s$?
\vs 0.3cm

Arguments have been advanced \cite{Zam1}\cite{Zam2} in support
of the conjecture that Green's functions of
\s$\phi_{22}\s$, for which exact expressions
in terms of finite-dimensional integrals
are known\footnote{\s the 4-point function, for example, is a bilinear
combination of the hypergeometric functions},
coincide with the scaling
limit of Green's functions of the field \s$\varphi\s$
in critical $\s P(\varphi)_2\s$ models,
with $\s P\s$ of degree $\s 2k+2\s $. For k=1,
one obtains this way the scaling limit
of Green's functions of the two-dimensional
critical Ising model with critical exponent
\s$\eta=4\Delta^{\pm}_{22}={1\over 4}\s$.
\s The model with \s$k=2\s$ should correspond to the tricritical
Ising model for which one obtains \s$\eta=4\Delta^\pm_{22}
={3\over{20}}\s$.
It is remarkable that,
although  detailed rigorous control of these scaling limits
has long eluded the attempts of constructive field theory, we
apparently have at our disposal the exact expressions for
the critical
exponents and even for limiting
Green's functions. It may be worthwhile to produce
a clean proof that one obtains this way an example of a theory
satisfying all (massless) quantum field theory axioms.
\vskip 0.8cm

\nsection{\hspace{-.7cm}.\ \ Supersymmetric CFT}
\vs 0.2cm
\no{\bf{4.1.\s\ \ Geodesic motion of a superparticle}}
\vs 0.4cm

In Sects.\s\s 1, 3.1 and 3.3, we have sketched how the
Riemannian geometry of a manifold \s$M\s$ is reflected
in the Laplace-Beltrami operator acting in \s$L^2(M)\s$.
For many purposes, however, especially for constructing
characteristic classes of the manifold and for
analyzing geometry with torsion, it is more
convenient to work with the Dirac operator \s$\Di\s$
acting in the space of sections of the spinor bundle
over \s$M\s$. A suitable framework may be obtained by
quantizing a supersymmetric (SUSY) version of
the geodesic motion on \s$M\s$.
Instead of trajectories \s$\tau\mapsto x(\tau)\s$,
one considers
\qq
{\bf x}^\mu(\tau,\theta^\pm)\equiv x^\mu(\tau)+\theta^+
\psi^{\mu+}(\tau)
+\theta^-\psi^{\mu-}(\tau)+\theta^+\theta^-F^\mu(\tau)\ ,
\qqq
where \s$x^\mu(\cdot)\s,\ \s F^\mu(\cdot)\s$ are
functions and \s$\psi^{\mu\pm}(\tau)\s$ as well as
\s$\theta^\pm\s$ are anticommuting Grassmann generators.
We could also consider a system with only one
\s$\theta\s$ parameter. The pair \s$\theta^\pm\s$ is
inherited from the left-right moving sectors
of the 1+1-dimensional SUSY sigma model if we restrict
it to fields constant in space, as explained below.
With the use of operators
\qq
D_\pm\equiv\da_{\theta^\pm}+i\theta^\pm\da_t\ ,
\qqq
the action functional for the geodesic
motion of a superparticle on
\s$M\s$ may be written as
\qq
S({\bf x}(\cdot))\s=\s\hf\int\gamma_{\mu\nu}
({\bf x})\s(D_+{\bf x}^\mu)
(D_-{\bf x}^\nu)\s\s d\tau\s d\theta^+ d\theta^-\ ,
\label{act}
\qqq
where the Berezin integration over \s$\theta^\pm\s$
is defined by the standard rule
\s$\int\hs{-0.04cm}d\theta^\pm=0\s,$
\s$\s\int\hs{-0.04cm}\theta^\pm d\theta^\pm=1\s$.
It will be convenient to add to the action(\ref{act})
the term
\qq
S'({\bf x}(\cdot))\s=\s\hf\int
\beta_{\mu\nu}({\bf x})\s(D_+{\bf x}^\mu)
(D_-{\bf x}^\nu)\s\s d\tau\s d\theta^+ d\theta^-
\qqq
with \s$\beta_{\mu\nu}\s dx^\mu\wedge dx^\nu\equiv\beta\s$
a 2-form on \s$M\s$.
Performing the \s$\theta^\pm\s$ integration (compare
Sect.\s\s23.1b of \cite{West}),
one obtains the
component expression for \s$S^{\rm tot}\equiv S+S'\s$:
\qq
\hs*{-0.5cm}S^{\rm tot}({\bf x}(\cdot))\s=
\s\int[\s\hf\s\gamma_{\mu\nu}
\s(\da_tx^\mu)(\da_tx^\nu)\s+\s{_i\over^2}\s
\gamma_{\mu\nu}\s\psi^{\mu+}\s\nabla^+_t\psi^{\nu+}
\s+\s{_i\over^2}\s
\gamma_{\mu\nu}\s\psi^{\mu-}\s\nabla^-_t\psi^{\nu-}\ \s\cr
+\s{_1\over^4}\s R^-_{\s\ \kappa\mu\lambda\nu}\s\psi^{\kappa+}
\psi^{\mu+}\psi^{\lambda-}\psi^{\nu-}\s-\s \gamma_{\rho\sigma}
\s(F^\rho-\Gamma^{-\s\rho}_{\ \ \ \s\mu\nu}\s\psi^{\mu+}\psi^{\nu-})
\s(F^\sigma-\Gamma^{-\s\sigma}_{\s\ \ \ \kappa
\lambda}\s\psi^{\kappa+}
\psi^{\lambda-})\s\s]\s\s\s d\tau\ ,
\label{action5}
\qqq
where
\qq
\nabla^\pm_t\psi^{\mu\pm}\s=\s\da_t\s\psi^{\mu\pm}\s+\s
\Gamma^{\pm\s\mu}_{\s\ \ \ \kappa\lambda}\s(\da_tx^\kappa)
\hs{0.03cm}\psi^{\lambda\pm}
\qqq
are the covariant derivatives with respect to connections
with torsion,
\qq
\Gamma^{\pm\s\mu}_{\s\ \ \ \kappa\lambda}&=
&\Gamma^\mu_{\ \kappa\lambda}\s\pm\hf\s\gamma^{\mu\nu}\s
H_{\nu\kappa\lambda}\ ,\label{TOR}
\qqq
\vs -0.3cm

\no where
\vs -0.3cm

\qq
\Gamma^\mu_{\ \kappa\lambda}\s=\s\hf\s\gamma^{\mu\nu}
(\da_\kappa\gamma_{\nu\lambda}+\da_\lambda\gamma_{\nu\kappa}
-\da_\nu\gamma_{\kappa\lambda})\ ,\hs{1cm}
H_{\nu\kappa\lambda}\s=\s\da_\nu\beta_{\kappa\lambda}+
\da_\kappa\beta_{\lambda\nu}+\da_\lambda\beta_{\nu\kappa}\ .
\qqq
$(\Gamma^\mu_{\ \kappa\lambda})\s$ define the Levi-Civita
connection, while \s$\Gamma^{\pm\s\mu}_{\s\ \ \ \kappa
\lambda}\s$ is modified by a torsion
term coming from the 2-form \s$\beta\s$.
\s Note that \s$\beta\s$ enters the above formulae only
through the torsion form \s$d\beta={1\over 3}H_{\nu\kappa\lambda}
\s dx^\nu\wedge
dx^\kappa\wedge dx^\lambda\s$.
\qq
R^\pm_{\s\ \kappa\mu\lambda\nu}\s=\s\da_\kappa
\Gamma^{\pm}_{\ \s\lambda\mu\nu}-\da_\mu\Gamma^{\pm}_{\s\
\lambda\kappa\nu}-\Gamma^\pm_{\s\ \rho\kappa\lambda}
\s\Gamma^{\pm\s\rho}_{\ \ \s\ \mu\nu}+\Gamma^\pm_{\s\ \rho\mu\lambda}
\s\Gamma^{\pm\s\rho}_{\ \ \s\ \kappa\nu}
\qqq
are the corresponding curvature tensors of the connections
with torsion\footnote{\s indices are
raised and lowered with the metric \s$\gamma\s$}.
In particular, for \s$\beta_{\mu\nu}=0\s$, they reduce to
the curvature of the Levi-Civita connection. Notice that
\s$F^\mu(\tau)\s$ are not dynamical variables. They
may be eliminated from their classical equations.
The classical supersymmetry is generated by the constants
of motion
\qq
Q^\pm\s\equiv\s{_1\over^{\sqrt{2}}}\s\s(\s
\mp\s\gamma_{\mu\nu}\s\da_t
x^\mu\psi^{\nu\pm}
\s+\s{_{i}\over^6}\s
H_{\mu\nu\kappa}\s\psi^{\mu\pm}\psi^{\nu\pm}
\psi^{\kappa\pm}\s)\ .
\label{susygen}
\qqq
\vs 0.3cm

It is not difficult to quantize the system. As the Hilbert
space \s$\NH\s$ of states, we shall take \s$L^2(S\otimes S)\s$ for
even dimensional \s$M\s$, or \s$L^2(S\otimes S\otimes\NC^2)\s$
for odd dimensional \s$M\s$, \s where \s$S\s$ is the bundle of
spinors (containing both chiralities in the even-dimensional case).
The fibers of \s$S\otimes S\s$ (\s$S\otimes S\otimes\NC^2\s$)
\s support the double Dirac algebra:
\qq
\{\Gamma^{A\pm},\s\Gamma^{B\pm}\}=-2\s\delta^{AB}\ ,\ \ \
\{\Gamma^{A\pm},\s\Gamma^{B\mp}\}=0\ ,\ \ \ (\Gamma^{A\pm})^*
=-\Gamma^{A\pm}
\label{Diral}
\qqq
related to a local vielbein\footnote{\s more exactly, to
its lift to the spin bundle covering twice the bundle
of vielbeine} \s$(e^\mu_A)\s$, \s$e^\mu_Ae^\nu_A
=\gamma^{\mu\nu}\s$, \s$\gamma_{\mu\nu}
e^\mu_Ae^\nu_B=\delta_{AB}\s$.
\qq
\psi^{\mu\pm}={_1\over^{i\sqrt{2}}}\s e^\mu_A\s\Gamma^{A\pm}
\qqq
will quantize the classical Grassmann generators
\s$\psi^{\mu\pm}(0)\s$.
\qq
\{\psi^{\mu\pm},\s\psi^{\nu\pm}\}=\s\gamma^{\mu\nu}\ ,\ \ \
\{\psi^{\mu\pm},\s\psi^{\nu\mp}\}=0\ ,\ \ \ (\psi^{\mu\pm})^*
=\psi^{\mu\pm}\ .
\qqq
Introducing the spin connections with torsion
by
\qq
\nabla^\pm_\nu e_B\equiv\omega^{\pm\s A}_{\nu\ \ B}\s e_A\ ,
\qqq
we shall define the left-right covariant derivative
of the double index spinors \s$\chi\s$ (sections of
\s$S\otimes S\s$ or of \s$S\otimes S\otimes \NC^2\s$) by
\qq
\nabla_\mu\chi\s=\s\da_\mu\chi\s-\s{_1\over^8}\sum\limits
_{+,-}\omega^{\pm\s A}_{\mu\ \ B}\s[\Gamma^{A\pm},
\s\Gamma^{B\pm}]\s\chi\ .
\label{COVDER}
\qqq
The classical supersymmetry generators \s$Q^\pm\s$ of
(\ref{susygen}) give rise, on the quantum level, to two
Dirac operators
\qq
\Di_\pm\s=\s{_1\over^{\sqrt{2}}}\s\s(\s
\pm\s i\s\psi^{\mu\pm}\nabla_\mu\s+
\s{_{i}\over^6}\s
H_{\mu\nu\kappa}\s\psi^{\mu\pm}\psi^{\nu\pm}
\psi^{\kappa\pm}\s)
\qqq
with the following simple (although tedious to verify) algebra:
\qq
\hf\CH\s\equiv\s\Di_+^2&=&\Di_-^2\s
=\s-{_1\over^4}\s\gamma^{\mu\nu}(\nabla_\mu
\nabla_\nu
-\Gamma^{\kappa}_{\ \mu\nu}\nabla_\kappa)\s+
{_1\over^{16}}\s R\s-\s{_1\over^{192}}\s H_{\kappa\mu\nu}
H^{\kappa\mu\nu}\cr
&&\hs{4.4cm}-\s{_1\over^{32}}\s R^-_{\ \mu\nu\kappa\lambda}
\s[\psi^{\mu+},\s\psi^{\nu+}]\s[\psi^{\kappa-},
\s\psi^{\lambda-}]\ ,\cr
\{\Di_+,\s\Di_-\}\ \ &=&\ 0\ ,
\qqq
where \s$R\s$ is the scalar curvature of the Levi-Civita
connection \s$(\Gamma^{\kappa}_{\ \mu\nu})\s$.
\vs 0.4cm
\vs 0.2cm
\no{\bf{4.2.\s\ \ Superparticle on a group}}
\vs 0.4cm

Among the simplest examples of superparticle motions is that
on the group manifold \s$G\s$, with the action
\qq
S^{\rm tot}(g,\psi^\pm)\s=\s-{_k\over^4}\int\tr\s(g^{-1}
\da_tg)^2\s d\tau\s+\s i\int
\tr\s(\psi^+\da_t\psi^+\s+\s\psi^-\da_t\psi^-)\s d\tau\ ,
\label{action6}
\qqq
where \s$\psi^\pm(\tau)=t^A\psi^{A\pm}(\tau)\s$ are the Lie
algebra \s${\bf g}$-valued Grassmann variables.
Let \s$e_A^\pm\s$ denote the right- (left-)invariant
vector fields on \s$G\s$ generated by \s${_2\over^{\sqrt{k}}}
t^A\in{\bf g}\s$. They yield two global vielbeine for a
left-right invariant Riemannian metric on \s$G\s$.
Setting \s$e_A^{\mu\pm}\psi^{A\pm}
\equiv\psi^{\mu\pm}\s$ relates
the variables \s$\psi^{A\pm}\s$ to \s$\psi^{\mu\pm}\s$
used above. The connections with torsion \s$\nabla^\pm\s$
preserve the vector fields \s$e_A^\pm\s$, \s respectively,
and, consequently, have no curvature. This explains
the simplicity of eq.\s\s(\ref{action6}) as compared
to (\ref{action5}). The torsion coefficients are given by
\qq
e_A^{\kappa\pm}\s e_B^{\mu\pm}
\s e_C^{\nu\pm}\s H_{\kappa\mu\nu}\s
=\s-{_2\over^{\sqrt{k}}}\s f^{ABC}
\qqq
(both signs in \s$e^\pm\s$ give the same expression)
and the torsion form \s${1\over 3}H_{\kappa\mu\nu}
\s dx^\kappa\wedge
dx^\mu\wedge dx^\nu={ik\over 3}\s\tr\s(g^{-1}dg)^{\wedge 3}\s$
and is only locally exact.
Using the global vielbeine \s$(e^\pm_A)\s$ to trivialize
the spinor bundles over \s$G\s$, we may identify
the Hilbert space of states with
\qq
\NH=L^2(G)\otimes W\ ,
\qqq
where \s$W\s$ carries an irreducible representation of
the double Dirac algebra (\ref{Diral}), with \s$\psi^{A\pm}=
{1\over i\sqrt{2}}\Gamma^{A\pm}\s$. \s By setting
\qq
j^{A\pm}&=&{_1\over^{2i}}\s f^{ABC}\s\psi^{B\pm}\psi^{C\pm}\ ,\cr
J^{A\pm}&=&\pm {_{i\sqrt{k}}\over^2}\nabla_{\hs{-0.03cm}e^\pm_A}\ ,
\qqq
we obtain four commuting representations
of the Lie algebra \s${\bf g}\s$ acting in \s$\NH\s$:
\qq
[j^{A\pm},\s j^{B\pm}]\s=\s i\s f^{ABC}\s j^{C\pm}\ ,\ \ \ \
[J^{A\pm},\s J^{B\pm}]\s=\s i\s f^{ABC}\s J^{C\pm}\ .
\qqq
The Dirac operators are
\qq
\Di_\pm\s=\s\sqrt{{_2\over^k}}\s(\psi^{A\pm} J^{A\pm}\s-
\s{_i\over^6}\s f^{ABC}
\psi^{A\pm}\psi^{B\pm}\psi^{C\pm})
\label{dir}
\qqq
and for the Hamiltonian one obtains
\qq
\hf\s\CH\s=\s\Di_\pm^2\s=\s{_1\over^k}\s J^{A\pm}J^{A\pm}\s+
\s{_{g^\vee{\rm dim}(G)}\over^{24\s k}}\ .
\qqq
Notice that \s$\Di_\pm^2\s$ is a strictly positive operator:
\qq
\Di_\pm^2\s\geq\s{_{g^\vee{\rm dim}(G)}\over^{24\s k}}\ .
\label{INEQ}
\qqq
\vs 0.4cm
\vs 0.2cm
\no{\bf{4.3.\s\ \ Elements of the non-commutative de Rham calculus}}
\vs 0.4cm

Suppose, for simplicity, that \s$M\s$ is compact.
Given the algebra \s$\CA=C^\infty(M)\s$ and a Dirac operator
\s$\Di=\Di_\pm\s$, one may rewrite the de Rham differential form
calculus, following Connes {\it et al.}
\cite{Connes0}\cite{ConnesLott}\cite{Connes1}, in terms
of that of operators
acting in the Hilbert space \s$H\s=\s L^2(S\otimes S)\s$
(or \s$L^2(S\otimes S\otimes\NC^2)\s$,
\s for \s$M\s$ odd-dimensional).
This reformulation carries over to the non-commutative
setting where \s$\Di\s$ is a selfadjoint
operator on a Hilbert space \s$H\s$ and \s$\CA\s$ is a
$*$-algebra of operators in \s$B(H)\s$ satisfying certain
regularity assumptions
\cite{Connes0}\cite{ConnesLott}\cite{Connes1}.
A brief exposition of these ideas\footnote{\s a different approach
to non-commutative de Rham
calculus may be found in \cite{DVKM}} is contained
in Sect.\s 6 of \cite{Connes2}, see
also \cite{ChFF}.
Consider operators \s$\omega\s$ acting in \s$H\s$
and given by the expressions
\qq
\omega\s=\s\sum\limits_ja_{j0}
[\Di, a_{j1}]\cdots[\Di, a_{j{n}}]\ ,
\label{difform}
\qqq
where \s$a_{jm}\in\CA\s$. \s The space
\s${\Omega'}^n(\CA)\s$ of such operators
clearly forms a left \s$\CA$-module.
Since for \s$a,\s b\in\CA\s$
$$[\Di\s, a]\s b=[\Di, ab]\s-\s a\s[\Di, b]\ ,$$
${\Omega'}^n(\CA)\s$ is also a right \s$\CA$-module
and \s$\Omega'(\CA)\equiv\oplus_n{\Omega'}^n(\CA)\s$
becomes a \s$\NZ$-graded algebra.
Let \s$N^n(\CA)\s$ be composed of operators
$$\eta\s=\s\sum\limits_j\s[\Di, a_{j0}]\s
[\Di, a_{j1}]\cdots[\Di, a_{j{n-1}}]$$
such that \s\s$\sum_ja_{j0}
[\Di,a_{j1}]\cdots[\Di,a_{j{n-1}}]=0\s.$ \s It is not
difficult to see that \s$N(\CA)=\oplus_nN^n(\CA)\s$
is a left-right \s$\NZ$-graded ideal
in \s$\Omega'(\CA)\s$.
Define the \s$\NZ$-graded algebra \s$\Omega(\CA)\equiv
\Omega'(\CA)/N(\CA)\s$.
Notice that \s$\Omega^0(\CA)={\Omega'}^0(\CA)=\CA\s$
and \s$\Omega^1(\CA)={\Omega'}^1(\CA)=\{\ \sum_ja_{j0}[\Di,a_{j1}]
\ \s|\ \s a_{j0},\s a_{j_1}\in \CA\ \}\s$.
Each \s$\Omega^n(\CA)\s$ is a left-right \s$\CA$-module.
On \s$\Omega(\CA)\s$, \s one may define the graded
differential \s$d\s$ acting on representatives by
\qq
d\s\sum\limits_ja_{j0}
[\Di, a_{j1}]\cdots[\Di, a_{j{n}}]\s=
\s\sum\limits_j\s[\Di,a_{j0}]
[\Di, a_{j1}]\cdots[\Di, a_{j{n}}]\ .
\qqq
\vs -0.2cm
\no We have
\vs -0.2cm
\qq
d(\omega\rho)=(d\omega)\s\rho+(-1)^{{\rm deg}(\omega)}
\s\omega\s d\rho\ .
\qqq
In the general setup, $\Omega^n(\CA)\s$ plays the role of
the space of smooth $n$-forms.
\vs 0.3cm

We shall be interested in the situations where the
trace of non-zero
operators in \s${\Omega'}^n(\CA)\s$ diverges,
but one may define its regularized version (which we shall
denote by the integral sign) for example by
\qq
\int\omega\equiv\lim_{\epsilon\searrow 0}
{\tr(\omega\s\ee^{-\epsilon\s {{\slash\hs{-0.15cm}\partial}}^2})\over
\tr(\ee^{-\epsilon\s {{\slash\hs{-0.15cm}\partial}}^2})}\ ,
\qqq
\vs -0.2cm
\no or using the more sophisticated Dixmier trace \cite{Connes1},
so that
\vs -0.2cm
\qq
(\omega,\rho)\equiv\int\omega^*\rho
\qqq
defines a scalar product on \s${\Omega'}^n(\CA)\s$.
We may then complete \s${\Omega'}^n(\CA)\s$ to a Hilbert
space and embed \s$\Omega^n(\CA)\s$ into the subspace
\s$\Lambda^n(\CA)\s$ of this completion
perpendicular to \s$N(\CA)\s$.
\s$\Lambda^n(\CA)\s$ plays the role of the space
of square integrable $n$-forms.
\vskip 0.3cm

In the commutative example, with \s$H=L^2(S\otimes S)\s$
\s$(\s L^2(S\otimes S\otimes \NC^2)\s)$\s, \s$\Di=\Di_\pm\s$
and \s$\CA=C^\infty(M)\s$,
$$\omega=(\pm i)^na_{j_0}(\da_{\mu_1}a_{j1})\cdots
(\da_{\mu_n}a_{jn})\s\psi^{\mu_1\pm}\cdots\psi^{\mu_n\pm}$$
($\pm$ depending on whether \s$\Di=\Di_+\s$ or
\s$\Di=\Di_-\s$)\s. \s The
class of \s$\omega\s$, represented by \s$\omega^\perp
\in{\Lambda^n}(\CA)\s$, \s may be identified
with the de Rham n-form
$$\tilde\omega=a_{j_0}(\da_{\mu_1}a_{j1})\cdots
(\da_{\mu_n}a_{jn})\s dx^{\mu_1}\hs{-0.05cm}
\wedge\dots\wedge dx^{\mu_n}\ .$$
The scalar product \s$(\omega^\perp,\rho^\perp)\s$
is then proportional to \s$\int_M*\tilde
\omega\wedge\tilde\rho\s$, \s where \s$*\tilde\omega\s$
is the Hodge star of \s$\tilde\omega\s$.
\vs 0.3cm

In the general setup, one defines a vector
bundle \s$E\s$ over \s$\CA\s$ as a finitely generated,
projec-\break tive\footnote{\s i.e. the image of an idempotent in a
finite free module} left \s$\CA$-module.
A connection on \s$E\s$ is defined as
a linear mapping \s$\nabla:E\rightarrow \Omega^1
(\CA)\otimes_{\CA}E\s$
such that, for \s$a\in\CA\s$ and \s$s\in E\s$,
\qq
\nabla(as)=da\otimes s+a\s\s \nabla s\ .
\label{connec}
\qqq
$\nabla\s$ may be uniquely extended to an endomorphism
of the graded left \s$\Omega(\CA)$-module \s$\Omega(\CA)
\otimes_{\CA}E\s$ satisfying
\qq
\nabla(\omega\phi)=(d\omega)\phi+(-1)^{{\rm deg}(\omega)}\omega\s\s
\nabla\phi\ .
\qqq
One may then define the curvature \s$R(\nabla)\s$ by
\qq
R(\nabla)=-\nabla^2|_E\ .
\label{curva}
\qqq
$R(\nabla): E\rightarrow \Omega^2(\CA)\otimes_{\CA}E\s$
and obeys \s$R(\nabla)(as)=a\s R(\nabla)s\s$ for \s$a\in\CA\s$
and \s$s\in E\s$, i.e. it is an \s$\CA$-tensor \cite{ChFF}.
\vs 0.3cm

If the left \s$\CA$-module \s$\Omega^1(\CA)\s$
is finitely generated and projective, we may view
it as the cotangent bundle. For a connection
\s$\nabla:\Omega^1(\CA)\rightarrow\Omega^1(\CA)\otimes_\CA
\Omega^1(\CA)\s$ one may define non-commutative
torsion \cite{ChFF} as the
map \s$T(\nabla):\Omega^1(\CA)
\rightarrow\Omega^2(\CA)\s$ given by
\qq
T(\nabla)=d-m\s\nabla\ ,
\label{TORNC}
\qqq
where \s$m:\Omega^1(\CA)\otimes_\CA\Omega^1(\CA)\rightarrow
\Omega^2(\CA)\s$ is the multiplication map. It follows
that \s$T(\nabla)(a\omega)=a\s T(\nabla)\omega\s\s$:
\s\s$T(\nabla)\s$ is an \s$\CA$-tensor.
Also the notions of Levi-Civita connection, Ricci curvature
and scalar curvature may be introduced in the non-commutative
setup \cite{ChFF}.
\vs 0.3cm

In the commutative case, the above definition of
(finite-dimensional) vector bundles coincides
with the standard one if we represent a vector bundle
by the module of its smooth sections.
In particular, commutative $\s\Omega^{1}(\CA)$ is a
projective, finitely
generated left(-right) \s$\CA$-module representing
the cotangent bundle and the covariant
derivative \s$\nabla_\mu\s$ given by (\ref{COVDER})
allows to define a linear map
\s\s$\nabla: \Omega^{1}(\CA)\rightarrow\Omega^1
\otimes_{\CA}\Omega^1(\CA)\s\s$ by
\qq
\nabla\omega=\sum\limits_ja_j\psi^\mu\otimes[\nabla_\mu,\omega]\ ,
\label{CON}
\qqq
where \s$(a_j)\s$ is a partition of unity subordinate to
a covering of \s$M\s$ by coordinate charts. \s$\psi^\mu=
\psi^{\mu\s+}\s$ or \s$\psi^{\mu\s-}\s$. \s The commutator
is that of the operators acting on sections of \s$S\otimes S\s$
(or of \s$S\otimes S\otimes\NC^2\s$)\s. Clearly, \s$\nabla\s$
satisfies (\ref{connec}) and represents the connections
with torsion given locally by (\ref{TOR}). Its curvature,
defined by eq.\s\s(\ref{curva}),
may easily be seen to correspond to the standard curvature
2-form. In particular, if \s$M=G\s$, \s
as in Sect.\s\s 4.2, then \s$\Omega^1(\CA)\s$
is a free \s$\CA$-module with a basis \s$(\psi^{A\s\pm})\s$.
Setting \s$\psi^\pm\equiv t^A\psi^{A\s\pm}\s$, we may identify
\s${\sqrt{2\over k}}\s\psi^+\s$ with the Maurer-Cartan
1-form \s$(dg)g^{-1}\s$ and \s${\sqrt{2\over k}}\s\psi^-\s$
with \s$g^{-1}dg\s$.
Eq.\s(\ref{CON}) reduces to
\qq
\nabla\omega={\sqrt{_2\over^k}}\s\s\psi^{A\s\pm}\otimes
[J^{A\s\pm},\s\omega]
\label{CON1}
\qqq
(recall that the sign depends on whether \s$\Di=\Di_+\s$ or
\s$\Di_-\s$)\s. \s Consequently, \s$\psi^{A\s\pm}\s$ are
parallel i.e. annihilated by \s$\nabla\s$
The corresponding
curvature tensor vanishes. We leave as an exercise
verification that the torsion \s$T(\nabla)\s$, \s as defined
by eq.\s\s(\ref{TORNC}) is given by the formula
\qq
T(\nabla)\psi^{A\s\pm}={_1\over^{2i}}\s{\sqrt{{_2\over^k}}}\s
f^{ABC}\psi^{B\s\pm}\psi^{C\s\pm}\ .
\qqq
\vs 0.3cm

{\un{Summarizing}}: the Riemannian geometry may
be encoded in the SUSY triple
\s$(\s L^2(S\otimes S(\otimes\NC^2))\s,
\s\Di_\pm\s,\s C^\infty(M)\s)\s$
from which one can recover
not only the metric \s$\gamma\s$
but also the de Rham calculus of differential forms
which, more conveniently, may be rewritten
in Connes' operator language;
we have seen how
the connections with the torsion 3-forms
\s$\pm{1\over 3}\s H_{\kappa\mu\nu}\s dx^\kappa
\wedge dx^\mu\wedge dx^\nu\s$
may be incorporated into the operator formalism of
\cite{Connes0}\cite{ConnesLott}\cite{Connes2} whose main
virtue is that it extends to non-commutative spaces
characterized by general unital \s$*$-algebra \s$\CA\s$.
\vskip 0.4cm
\vs 0.2cm
\no{\bf{4.4.\s\ \ SUSY WZW model}}
\vs 0.4cm

As we have seen above, the natural generalizations
of the geodesic motion of a particle on a Riemannian
manifold \s$M\s$ to the case of 1+1-dimensional field
theory is given by the sigma model with \s$M\s$ as the
target. Similarly, the geodesic motion of a superparticle
generalizes to an (N=1) SUSY sigma model
with fields \s${\bf x}(\sigma,\tau,\theta^\pm)\s$
and an action \cite{West}
\qq
S^{\rm tot}({\bf x}(\cdot))\s=\s{_1\over^{4\pi}}\int
(\gamma_{\mu\nu}+\beta_{\mu\nu})({\bf x})\s(D_+{\bf x}^\mu)
(D_-{\bf x}^\nu)\s\s d\sigma\s d\tau\s d\theta^+d\theta^-\ ,
\qqq
where now
\qq
D_\pm\equiv\da_{\theta^\pm}+i\theta^\pm\da_\pm\ .
\qqq
with the light-cone derivatives
\s$\da_\pm\equiv\da_\tau\pm\da_\sigma\s$. \s
In components,
\qq
\hs*{-0.4cm}S^{\rm tot}({\bf x}(\cdot))\s=\s{_1\over^{2\pi}}\int
[\s\hf\s(\gamma_{\mu\nu}+
\beta_{\mu\nu})\s(\da_+x^\mu)(\da_-x^\nu)\s+\s{_i\over^2}\s
\gamma_{\mu\nu}\s\psi^{\mu+}\s\nabla^+_-\psi^{\nu+}
\s+\s{_i\over^2}\s
\gamma_{\mu\nu}\s\psi^{\mu-}\s\nabla^-_+\psi^{\nu-}\s\cr
\hs*{-0.55cm}+\s{_1\over^4}\s R^-_{\s\ \kappa
\mu\lambda\nu}\s\psi^{\kappa+}
\psi^{\mu+}\psi^{\lambda-}\psi^{\nu-}-\s \gamma_{\rho\sigma}
\s(F^\rho-\Gamma^{-\s\rho}_{\ \ \ \s\mu\nu}
\s\psi^{\mu+}\psi^{\nu-})
\s(F^\sigma-\Gamma^{-\s\sigma}_{\s\ \ \
\kappa\lambda}\s\psi^{\kappa+}
\psi^{\lambda-})\s\s]\s\s\s d\tau\ ,\ \s
\qqq
where
\qq
\nabla^\pm_+\psi^{\mu\pm}\s=\s\da_+\psi^{\mu\pm}\s+\s
\Gamma^{\pm\s\mu}_{\s\ \ \ \kappa\lambda}\s(\da_+x^\kappa)
\psi^{\lambda\pm}\ ,\hs{0.02cm}
\nonumber
\qqq
\vskip -1cm
\qq
\nabla^\pm_-\psi^{\mu\pm}\s=\s\da_-\psi^{\mu\pm}\s+\s
\Gamma^{\pm\s\mu}_{\s\ \ \ \kappa\lambda}(\da_-x^\kappa)
\psi^{\lambda\pm}\ \ \s
\nonumber
\qqq
and the supersymmetry generators
\qq
Q^\pm\s\equiv\s{_1\over^{\sqrt{2}}}\s\s(\s\mp\s
\gamma_{\mu\nu}\s\da_\pm x^\mu\psi^{\nu\pm}
\s+\s{_{i}\over^6}\s
H_{\mu\nu\kappa}\s\psi^{\mu\pm}\psi^{\nu\pm}
\psi^{\kappa\pm}\s)
\label{susygen1}
\qqq
satisfy \s\s$\da_\mp Q^\pm\s=\s0\ .$
\vs 0.3cm

On the quantum level, the SUSY sigma model
still requires renormalizations, although somewhat less
severe ones than the purely bosonic model. As in the bosonic case,
instead of attempting a direct construction of the models,
one can use symmetry principles to obtain a rich family
of exactly solvable SUSY CFT's. The simplest
of them is the SUSY WZW model. It corresponds
to the action
\qq
S^{\rm tot}(g(\cdot),\s\psi^\pm(\cdot))\s=\s
S_{_{\rm bos}}(g)\s+\s{_i\over^{2\pi}}\int\tr\s\s(\psi^+
\da_-\psi^++\psi^-\da_+\psi^-)\s d\sigma\s d\tau\ ,
\qqq
where \s$S_{_{\rm bos}}(g)\s$ is the action of the bosonic
WZW model with group \s$G\s$ and level \s$k-g^\vee\s$.
The space of states
\s$\NH\s$ of the SUSY WZW model is the tensor product of
of the space of states \s$\NH_{_{\rm bos}}\s$
of the bosonic model, discussed before, (with the shifted \s$k\s$)
\s and of the Fock space \s$\CF\s$
of free Majorana-Fermi field \s$\psi^\pm(z^\pm)
\equiv t^A\psi^{A\pm}(z^\pm)\s$
with values in the Lie algebra \s${\bf g}\s$.
\qq
\psi^{A\pm}(z_\pm)=\sum\limits_nz_\pm^{-n-1/2}\s\psi^{A\pm}_{n}\ ,
\qqq
where the sum runs over \s$n\in\NZ+{1\over 2}\s$
in the Neveu-Schwarz sector and over \s$n\in\NZ\s$
in the Ramond sector (corresponding to two choices
of the spin structure on \s$S^1\s$)\s.
\qq
\{\s\psi^{A\pm}_{n}\s,\s\psi^{B\pm}_{m}\}\s=\s\delta^{AB}\s
\delta_{n+m,0}\ ,
\ \ \ \ \{\s\psi^{A+}_{n}\s,\s\psi^{B-}_{m}\}\s=\s0\ ,\ \ \ \
(\psi^{A\pm}_{n})^*=\psi^{A\pm}_{-n}\ .
\qqq
The Neveu-Schwarz sector
Fock space \s$\CF^{\rm NS}\s$ is obtained by applying
to the vacuum state \s$|{\rm vac}\rangle\s$ annihilated
by \s$\psi^{A\pm}_{n}\s$,
$n>0\s$, \s polynomials in \s$\psi^{A\pm}_{n}\s,
\ n<0\s$ (\s$n\s$ half-integer). \s The Ramond sector
Fock space \s$\CF^{R}\s$ arises by applying
polynomials in \s$\psi^{A\pm}_{n}\s,\ n<0\s,$ \s to the
vector space \s$W\s$ carrying the irreducible representation
of the \s$\psi^{A\pm}_{0}\s$ Clifford algebra and annihilated
by \s$\psi^{A\pm}_{n}\s,\ n>0\s$ (\s$n\s$ integer).
The total fermionic Fock space of the model is
\qq
\CF\s=\s\CF^{\rm NS}\oplus\CF^{\rm R}\ .
\qqq
It carries a representation of two commuting \s$\hat{\bf g}\s$
current algebras given by
\qq
j^{A\pm}_n\s=\s{_1\over^{2i}}\s f^{ABC}\sum\limits_m:
\psi^{B\pm}_{m}\psi^{C\pm}_{n-m}:
\qqq
and of the Virasoro algebras
\qq
l^\pm_n\s=\s-\sum\limits_{m}\hf m:\psi^{A\pm}_{m}\psi^{A\pm}_{n-m}:
\ \s(\s+\s{_{3\s{\rm dim}(G)}\over^{48}}\delta_{n,0}\s)
\qqq
(the constant should be added in the Ramond sector).
The complete space of states
\qq
\NH\s\equiv\s\NH^{\rm NS}\oplus\NH^{\rm R}\s=\s(\NH_{_{\rm bos}}
\otimes\CF^{\rm NS})\oplus(\NH_{_{\rm bos}}\otimes\CF^{\rm R})
\qqq
carries the representation of the left-right
Virasoro algebras with generators
\qq
L^\pm_n\s=\s L^\pm_{_{\rm bos\s}n}\s+\s l_n^\pm\ ,
\qqq
where \s$L^\pm_{_{\rm bos\s}n}\s$ are the Sugawara generators
in \s$\NH_{_{\rm bos}}\s$ constructed from the bosonic currents
\s$J^{A\pm}_n\s$.
\vs 0.3cm

The total Virasoro algebras \s$(L^\pm_n)\s$, which
have central charge
\s$c^G_{_{\rm SUSY\s}k}$ $=$ ${(k-g^\vee)\s{\rm dim}(G)\over k}\s+
\s{{\rm dim}(G)\over 2}\s$, \s
may be extended to the super-Virasoro
ones with additional generators
\qq
Q^\pm_n\s=\s\sqrt{{_2\over^k}}\s\s(\s\sum\limits_{m}
\psi^{A\pm}_{m}J^A_{n-m}\s-\s{_i\over^6}\s f^{ABC}
\sum\limits_{m,r}:\psi^{A\pm}_{m}\psi^{B\pm}_{r}
\psi^{C\pm}_{n-m-r}:\s)
\qqq
satisfying
\qq
[L^\pm_n\s,\s Q^\pm_m]\ &=&({_n\over^2}-m)\s Q^\pm_{n+m}\ ,\cr
\{Q^\pm_n\s,\s Q^\pm_m\}&=&2\s L^{\pm}_{n+m}\s+\s{_1\over^3}
\s c(n^2-{_1\over^4})\s\delta_{n+m,0}\ ,
\qqq
where the operators \s$Q^\pm_n\s$ commute with
\s$L^\mp_m\s$ and anticommute
with \s$Q^\mp_m\s$. In particular,
\qq
(Q^\pm_0)^2\s=\s L^\pm_0\s-\s{_c\over^{24}}\hs{1cm}{\rm and}
\hs{1cm}\{Q^+_0\s,\s Q^-_0\}=0
\label{N=1SUS}
\qqq
in the Ramond sector. Notice that the commutation relations
(\ref{N=1SUS}) may be rewritten as the global
SUSY algebra
\qq
\{Q_\alpha,\s Q_\beta\}\s=
\s (\gamma^\mu C)_{\alpha\beta}\s P_\mu\ ,
\label{N=1SUSY}
\qqq
where \s\s$\gamma^0=C\s=\left(\matrix{0&i\cr-i&0}\right)\s$,
\ \s$\gamma^1\s=\left(\matrix{0&i\cr i&0}
\right)\s$ are two-dimensional Dirac matrices and
\s\s$Q_1\equiv Q^+_0\s$, \s\s$Q_2\equiv Q^-_0\s$,
\s\s$P_0\equiv\CH=L^+_0+L^-_0-{1\over12}c^G_{{}_{\rm SUSY\s}k}\s$,
\s\s$P_1\equiv \CP=L^+_0-L^-_0\s$. The bottom of the
spectrum of the operators \s$Q_\alpha^2\s$ is attained on
states \s$|{\rm vac}\rangle_{_{\rm bos}}\otimes w\in\NH^{\rm R}\s$,
\s where \s$|{\rm vac}\rangle_{_{\rm bos}}\s$ is the
bosonic vacuum in \s$\NH_{_{\rm bos}}\s$ and \s$w\in W\subset
\CF^{\rm R}\s$. It is equal to \s${g^\vee\s{\rm dim}(G)
\over 24\s k}>0\s$ (compare to (\ref{INEQ})\s)\s. \s It follows
that there are no Ramond ground states (states annihilated
by \s$Q^\pm_0\s$) in the SUSY WZW model. Hence the global
supersymmetry (\ref{N=1SUSY}) is broken \cite{WittSUSY} and
the Witten index \s${\rm Tr}\s(-1)^F\s$, \s to which
only the Ramond
ground states contribute, vanishes.
\vs 0.4cm

The geometry of the SUSY WZW CFT may be encoded in the triple
\qq
(\NH^{\rm R},\s Q^\pm_0,\s \CA)\ ,
\label{TRIP}
\qqq
where \s$\CA\s$ is the bosonic algebra generated by
the primary fields of the bosonic WZW model (acting trivially
on the fermionic Fock space \s$\CF\s$)\s. \s The more general
field algebra (of bosonic and fermionic operators)
may be obtained by considering non-commutative differential
forms
\qq
\sum\limits_ja_{j0}\s\hs{0.03cm}[Q^\pm_0,\s a_{j1}]
\cdots[Q^\pm_0,\s a_{j{n_j}}]\ ,
\qqq
where \s$a_{jm}\in \CA\s$, \s compare with eq.\s\s(\ref{difform}).
In particular, operators \s\s$\sqrt{{_2\over^k}}\sum_ja_{j0}
\sum_n[J^{A\s\pm}_{-n},\s a_{j1}]\s\psi^{A\s\pm}_{n}\s\s$
are 1-forms.
\vs 0.3cm

The non-commutative geometry of the triple (\ref{TRIP})
is clearly infinite-dimensional and its treatment
would require serious analysis. This may be avoided
by restriction to the effective target geometry
of the SUSY WZW model which one may define
as follows. Let
\qq
\NH^{{\rm NS},{\rm R}}_0\s=\s\{\s\s|\phi\rangle\in
\NH^{{\rm NS},{\rm R}}
\ \ |\ \ \psi^{A\pm}_{n}|\phi\rangle
=0=J^{A\pm}_n|\phi\rangle
\ \ {\rm for}\ \s n>0\s\s\}
\qqq
be the small spaces of states. Notice that
\qq
\NH^{\rm NS}_0\s\cong\s\NH_{_{\rm bos} 0}\s\subset\s
L^2(G)\ ,\ \ \ \ \ \ \NH^{\rm R}_0\s\cong\s\NH_{_{\rm bos} 0}
\otimes W\s\subset\s L^2(G)\otimes W\ ,
\qqq
where \s$\NH_{_{\rm bos} 0}\cong L^2_{k-g^\vee}(G)\s$
is the small space
of the bosonic model obtained by restricting the sum in
(\ref{decomp}) to the representations integrable at level
\s$k-g^\vee\s$.
\s Operators \s$Q_0^\pm\s$ preserve \s$\NH^{\rm R}_0\s$
and reduce on it
to the Dirac operators \s$\Di_\pm\s$ of (\ref{dir}).
As for the algebra \s$\CA_0\s$, we shall choose it to be
the small bosonic algebra acting trivially on the \s$W\s$
factor of \s$\NH_0\s$\s. It is generated by the primary fields
corresponding to states \s$|\phi\rangle\in\NH^{\rm NS}_0\s$
sandwiched between the projectors on \s$\NH^{\rm R}_0\s$.
\s Clearly, the finite
non-commutative geometry encoded by \s$(\NH^{\rm R}_0,\s
Q^\pm_0|_{\NH^{\rm R}_0}\s,
\s \CA_0)\s$ is a deformation of the geodesic motion of the
superparticle on \s$G\s$ to which it reduces in the limit
\s$k\rightarrow\infty\s$.
\vs 0.3cm

For \s$(\s\NH_0\s,\s Q^\pm_0|_{\NH_0}\s,\s\CA_0\s)\s$,
\s (non-commutative)
1-forms are \s\s$\sqrt{{2\over^k}}\sum_ja_{j0}[J^{A\s\pm}_0,\s
a_{j_1}]\s\psi^{A\s\pm}_0\s\s$, \s with \s$a_{j0},$\s
$a_{j1}$\s${}\in\CA_0\s$, \s and the question arises whether any
operator \s$a^A\psi^{A\s\pm}_0\s$,  \s$a^A\in\CA_0\s$,
\s may be cast in such a form. In other words, is the
\s$\CA_0$-module \s$\Omega^1(\CA_0)\s$ free, with a
basis given by \s$(\psi^{A\s\pm}_0)\s$? \s This appears
to be a technically rather difficult question. The answer is yes for
\s$G=SU(2)\s$ and \s$k-g^\vee=k-2\leq 2\s$. If it is positive
in general then one may define
a connection \s$\nabla\s$ on \s$\Omega^1(\CA_0)\s$
essentially by the same formula (\ref{CON})
as used in the commutative case of SUSY quantum mechanics
on \s$G\s$:
\qq
\nabla\omega={\sqrt{_2\over^k}}\s\psi^{A\s\pm}_0\otimes
[J^{A\s\pm}_0,\s\omega]\ .
\qqq
Just as there, \s$\psi^{A\s\pm}_0\s$ would provide
a parallel basis of \s$\Omega^1(\CA_0)\s$
and \s$\nabla\s$ would have vanishing curvature:
it appears that the effective target of the SUSY WZW model
is a non-commutative space supplied with flat connections
of non-zero torsion. Although this has been proven only in the limit
\s$k=\infty\s$ and for \s$G=SU(2)\s$ at \s$k-2=1,2\s$, it is
reasonable to expect that it is true more generally.
\vs 0.4cm
\vs 0.2cm
\no{\bf{4.5.\ \ \s SUSY coset mechanics}}
\vs 0.4cm

Let us return to the motion of a superparticle
on a group \s$G\s$, \s with \s$L^2(G)\otimes W\s$ giving
the space of quantum states. Let \s$H\s$ be a connected
subgroup of \s$G\s$. We shall extend the
construction of the coset quantum mechanics to the SUSY case.
The generators \s$J^{\rm a\pm}\s$ of left-right regular actions
of \s${\bf h}\subset{\bf g}\s$ on \s$L^2(G)\s$ define two
commuting representations\footnote{\s recall that
\s${\rm a,b,}\dots\s$ label the generators of \s${\bf h}
\subset{\bf g}\s$ and \s$\alpha,\beta,\dots\s$
the ones in the perpendicular subspace \s${{{\bf h}^\perp}}\s$}
of \s$\bf h\s$ in \s$L^2(G)\otimes W\s$.
By taking
\qq
\tilde J^{{\rm a}\pm}\s\equiv\s J^{{\rm a}\pm}\s
+\s{_1\over^{2i}}\s f^{{\rm a}\beta\gamma}
\s\psi^{\beta\pm}\psi^{\gamma\pm}\ ,
\qqq
one obtains another such pair of representations.
As the space of states of the coset quantum
mechanics, we shall take
\qq
(L^2(G)\otimes W)_H\s\equiv\s\{\ |\phi\rangle\in
L^2(G)\otimes W\ \ |\ \ (\psi^{{\rm a}+}+\psi^{{\rm a}-})\s
|\phi\rangle=0=(\tilde J^{{\rm a}+}
+\tilde J^{{\rm a}-})\s|\phi\rangle\ \}\ .
\qqq
Let
\qq
\Di\hs{0.03cm}'\hs{-0.13cm}_\pm\s
=\s\sqrt{{_2\over^k}}\s\s(\psi^{{\rm a}\pm}
{\tilde J}^{{\rm a}\pm}
-\s{_i\over^6}\s f^{{\rm a}{\rm b}{\rm c}}
\psi^{{\rm a}\pm}\psi^{{\rm b}\pm}
\psi^{{\rm c}\pm}\s)\ .
\qqq
Then
\qq
\Di^{\rm cs}_{\pm}\equiv\Di_\pm
-\Di\hs{0.03cm}'\hs{-0.13cm}_\pm\s=\s
\sqrt{{_2\over^k}}\s\s(\psi^{\alpha\pm}
J^{\alpha\pm}
-\s{_i\over^6}\s f^{\alpha\beta\gamma}
\psi^{{\alpha}\pm}\psi^{{\beta}\pm}
\psi^{{\gamma}\pm}\s)
\qqq
commutes with \s$\psi^{{\rm a}\pm}\s$ and with \s$\tilde
J^{{\rm a}\pm}\s$ and defines the coset Dirac operators
acting on \s$(L^2(G)\otimes W)_H\s$. Finally, the natural
action of \s$C^{\infty}(G)_H\s$ on \s$L^2(G)\otimes W\s$
descends to \s$(L^2(G)\otimes W)_H\s$, \s so that we
may regard
\qq
(\s(L^2(G)\otimes W)_H\s,\s\Di^{\rm cs}_\pm,\s C^\infty(G)_H\s)
\label{CSM}
\qqq
as the triple encoding the geometry of the
SUSY coset quantum mechanics.
\vs 0.3cm

Let
\qq
W^\perp\s=\s\{\ |\phi\rangle\in W\ \s|\ \s
(\psi^{{\rm a}+}+\psi^{{\rm a}-})\s|\phi\rangle=0\ \}\ .
\qqq
$W^\perp\s$ carries the (irreducible) representation
of the algebra generated by \s$\psi^{\alpha\pm}\s$.
Note that the generators \s$\tilde J^{{\rm a}\pm}\s$
of \s$\bf h\s$ may be restricted to
\s$L^2(G)\otimes W^\perp\s$. \s
\s$(L^2(G)\otimes W)_H\s$ is the subspace
of \s$L^2(G)\otimes W^\perp\s$ invariant under
the diagonal action of \s$H\s$ defined by
\s$(\tilde J^{{\rm a}+}+\tilde J^{{\rm a}-})\s$. This picture
of SUSY coset quantum mechanics may be easily obtained
by a quantization of the classical system given by
the action functional
\qq
S(g(\cdot),\psi^{\alpha\pm}(\cdot),
A_\pm(\cdot))\s=\s-{_k\over^4}\int\tr\s\s(g^{-1}
\da_\tau g)^2\s d\tau\hs{5.5cm}\cr
+\s{_k\over^2}\int\tr\s\s(\s(g\s\da_\tau g^{-1})
A_-+A_+ (g^{-1}\da_\tau g)+gA_+
g^{-1}A_--A_+A_-\s)\s\s d\tau\cr
+\s{i\over^2}\sum\limits_{+,-}
\int\psi^{\beta\pm}(\delta^{\beta\gamma}
\da_t-if^{{\rm a}\beta\gamma}A_\mp^{\rm a})\s
\psi^{\gamma\pm}\s d\tau
\qqq
with the Grassmann variables  \s$\psi^{\alpha\pm}(\tau)\s$
solely in the \s${{\bf h}^\perp}\s$ directions and gauge fields
\s$A_\pm\equiv t^{\rm a}A^{\rm a}_\pm\s$
with values in \s${\bf h}\s$. Elimination of \s$A_\pm\s$
from the action leads to the action for a superparticle
moving on \s$G/{\rm Ad}(G)\s$ with the metric given by
eqs.\s\s (\ref{MTR}) or (\ref{MTRD}).
Recall that the (co-)vectors tangent to
\s$G/{\rm Ad}(H)\s$ could be
parametrized by elements of \s${\bf h}^\perp\s$
by formulae (\ref{bASER}) or (\ref{bASER}). These
are the parametrizations which allow to view
\s$t^\alpha\psi^{\alpha\s\pm}(\tau)\s$ as taking values in
the spaces tangent to \s$G/{\rm Ad}(H)\s$. It is
possible then to recover the torsion form from
the effective action obtained by eliminating
the gauge fields. It is given by the expression
\qq
{_{ik}\over^{3}}\s\tr\s(g^{-1}dg)^{\wedge 3}\s+\s{{ik}}\s
d\s\s\tr\s\s(g^{-1}dg)\s(E-E\s{\rm Ad}_{g}\s E)^{-1}
(g\s dg^{-1})
\qqq
which defines a, possibly singular,
closed 3-form on \s$G/{\rm Ad}(H)\s$.
\vs 0.3cm

Because of the presence of dilaton,
the coset quantum mechanics as defined by the triple
(\ref{CSM}) does not, however, coincide with
the one obtained by quantization of
the superparticle on \s$G/{\rm Ad}(H)\s$
with the metric and torsion form described above.
It differs from those by ordering effects.
They are rather complicated but the non-commutative
formalism offers a possibility to study
the coset geometry directly, using data (\ref{CSM}).
It is not hard to see, for example, that 1-forms
\s$\omega=\sum_ja_{j0}[\Di_\pm,a_{j1}]\s$ are arbitrary
operators on \s$(L^2(G)\otimes W)_H\s$ of the form
\s$a^\alpha\psi^{\alpha\s\pm}\s$, \s where \s$a^\alpha\s$
are functions on \s$G\s$ such that
\qq
[J^{{\rm a}+}+J^{{\rm a}-},\s a^\alpha]=
if^{{\rm a}\alpha\beta}a^\beta\ .
\qqq
We may identify \s$\psi^{\alpha\s\pm}\s$ with the standard
1-forms on \s$G\s$ given by eqs.\s\s (\ref{bASEL}) and
(\ref{bASER}). Formula
\qq
\nabla\omega={\sqrt{_2\over^k}}\s\psi^{\alpha\s\pm}\otimes
[J^{\alpha\s\pm},\s\omega]
\qqq
defines again a connection on the cotangent bundle
\s$\Omega^1(\CA)\s$ which, however, has non-trivial
curvature (also the Ricci and the scalar ones) and
torsion.
\vs 0.4cm
\vs 0.2cm
\no{\bf{4.6.\ \ \s SUSY coset CFT}}
\vs 0.4cm

The SUSY coset construction generalizes to the supersymmetric
case \cite{GKO}\cite{KacTod}. Introducing in
the Hilbert space of the SUSY WZW model
based on the group \s$G\s$ the modified
\s$\hat{\bf h}\subset\hat{\bf g}\s$ currents
\qq
\tilde J^{{\rm a}\pm}_n\s=\s J^{{\rm a}\pm}_n\s+\s{_1\over^{2i}}
\s f^{{\rm a}\beta\gamma}\sum\limits_m
:\psi^{\beta\pm}_m\psi^{\gamma\pm}_{n-m}:
\qqq
with central charge \s$k-h^\vee\s$, we may define the Hilbert
space of the coset theory as
\qq
\NH\equiv\NH^{\rm NS}\oplus\NH^{\rm R}
\s=\s\{\ \s\s|\phi\rangle\in\NH_{_{\rm WZW}}\ \ |\ \ \s
{\hbox to 4.6cm{${}_{\psi^{{\rm a}\pm}_n|\phi\rangle=0
=\tilde J^{{\rm a}\pm}_n|\phi\rangle\ \ {\rm for}\ \s n>0\ ,}$\hfill}
\atop\hbox to 4.6cm{${}^{(\psi^{{\rm a}+}_0
+\psi^{{\rm a}-}_0)|\phi\rangle=0
=(\tilde J^{{\rm a}+}_0+\tilde J^{{\rm a}-}_0)
|\phi\rangle}$\hfill}}\ \}\ ,
\qqq
where \s$\NH_{_{\rm WZW}}\s$ now denotes the space of states of the
SUSY WZW model with group \s$G\s$. \s
The operators
\qq
L^{{\rm cs\s}\pm}_n\s=\s L^\pm_{_{\rm bos\s}n}\s+\s l^\pm_n
-{L'}^\pm_{_{\rm bos\s}n}\s-\s {l'}^\pm_n\ ,
\qqq
where
\qq
L^\pm_{_{\rm bos\s}n}\s\s\s=&{_1\over^k}\sum\limits_m:J^{A\pm}_m\s
J^{A\pm}_{n-m}:\ ,\ \ \ \ l^\pm_n\s\s=&-\sum\limits_{m}
\hf m:\psi^{A\pm}_{m}\psi^{A\pm}_{n-m}:
\ \s(\s+\s{_{3\s{\rm dim}(G)}\over^{48}}\delta_{n,0}\s)\ ,\cr
{L'}^\pm_{_{\rm bos\s}n}\s=&{_1\over^k}\sum\limits_m:\tilde
J^{{\rm a}\pm}_m\s\tilde J^{{\rm a}\pm}_{n-m}:\ ,\ \ \ \
{l'}^\pm_n\s=&-\sum\limits_{m}
\hf m:\psi^{{\rm a}\pm}_{m}\psi^{{\rm a}\pm}_{n-m}:
\ \s(\s+\s{_{3\s{\rm dim}(H)}\over^{48}}\delta_{n,0}\s)\ \s
\qqq
preserve \s$\NH\s$ and define on it an action
of a commuting pair of Virasoro algebras with central charge
\s\s$c^G_{_{\rm SUSY\s}k}-c^H_{_{\rm SUSY\s}k}\s$.
\s Let
\qq
{Q'}^\pm_n\s=\s\sqrt{{_2\over^k}}\s\s(\s\sum\limits_{m}
\psi^{{\rm a}\pm}_{m}\tilde J^{{\rm a}\pm}_{n-m}\s
-\s{_i\over^6}
\s f^{{\rm a}{\rm b}{\rm c}}\sum\limits_{m,r}
:\psi^{{\rm a}\pm}_{m}\psi^{{\rm b}\pm}_{r}
\psi^{{\rm c}\pm}_{n-m-r}:\s)\ .
\qqq
Then
\qq
Q^{\rm cs\s\pm}_n\equiv Q^\pm_n-{Q'}^\pm_n\s=
\s\sqrt{{_2\over^k}}\s\s(\s\sum\limits_{m}
\psi^{{\alpha}\pm}_{m} J^{{\alpha}\pm}_{n-m}\s
-\s{_i\over^6}
\s f^{{\alpha}{\beta}{\gamma}}\sum\limits_{m,r}
:\psi^{{\alpha}\pm}_{m}\psi^{{\beta}\pm}_{r}
\psi^{{\gamma}\pm}_{n-m-r}:\s)
\qqq
preserve \s$\NH\s$ and extend the Virasoro algebras
\s$(L^{\rm cs\s\pm}_n)\s$ to the super-Virasoro ones.
\vs 0.3cm
On the Lagrangian level, the SUSY coset models
correspond to the action
\qq
S(g,\s\psi^{\alpha\pm},A_\pm)\s=\s S_{_{\rm bos}}(g,A_\pm)
\s+\s{_i\over^{4\pi}}\sum\limits_{+,-}\int\psi^{\beta\pm}
(\delta^{\beta\gamma}\da_\mp-if^{{\rm a}\beta\gamma}
A^{\rm a}_\pm)\s
\psi^{\gamma\pm}\s d\sigma\s d\tau\ ,
\qqq
where \s$S_{_{\rm bos}}(g,A_\pm)\s$ is as in
(\ref{gauact}), but with \s$k\mapsto k-g^\vee\s$,
\s$\psi^\pm\s$ take values in \s${\bf h}^\perp\s$
and \s$A_\pm\s$ in \s${\bf h}\s$.
\vs 0.3cm

Let \s$\NH^{\rm NS}_0\s$ (\s$\NH^{\rm R}_0\s$)
denote the subspace
in the Neveu-Schwarz (Ramond) sector of the coset
theory annihilated
by \s$\psi^{A\pm}_n\s$ and \s$J^{A\pm}_n\s$, \s
for \s$n>0\s$. Notice that \s$\NH^{\rm NS}_0\s$ may be
naturally identified with \s$L^2_{k-g^\vee}(G)_H\s$,
the subspace of states of the bosonic coset quantum mechanics
corresponding to the integrable representations.
Similarly
\s$\NH^{\rm R}_0\s$ is naturally isomorphic to
\s$(L^2_{k-g^\vee}(G)\otimes W)_H\s$ i.e. to a subspace of
the space of states of
the SUSY coset quantum mechanics. The action of
\s$Q^{\rm cs\s\pm}_0\s$ on \s$\NH^{\rm R}_0\s$ reduces to
that of \s$\Di^{\rm cs}_\pm\s$.
The (super-Virasoro) primary fields \s$V^{\rm cs}_{|\phi
\rangle}(z_\pm)\s$ of the
coset theory corresponding to vectors \s$|\phi\rangle\in
\NH^{\rm NS}\s$ annihilated by \s$Q^{{\rm cs}\s\pm}_n\s$
and \s$L^{{\rm cs}\s\pm}_n\s$ for \s$n>0\s$
may be used to generate the bosonic field
algebra \s$\CA\s$. Similarly, operators
\s\s$E^{\rm R}_0\s V^{\rm cs}_{|\phi\rangle}(1)\s
|_{\NH^{\rm R}_0}\s$ with \s$|\phi\rangle\in\NH^{\rm NS}_0\s$,
\s where \s$E^{\rm R}_0\s$ is the
orthogonal projection on \s$\NH^{\rm R}_0\s$, \s generate
a small algebra \s$\CA_0\s$. The triple \s\s$(\NH^{\rm R}\s,\s
Q^{\rm cs\s\pm}_0\s,\s \CA)\s$
represents the geometry of the SUSY coset CFT,
whereas \s$(\NH^{\rm R}_0\s,\s Q^{\rm cs\s\pm}_0
|_{{\NH}^{\rm R}_0}\s,\s\CA_0)\s$
encodes that of its effective target. The
latter is a finite non-commutative
deformation of the geometry of the SUSY coset quantum
mechanics discussed above.
\vs 0.3cm

As in the bosonic case, one may alternatively define an effective
target geometry of the SUSY coset models
using the SUSY extension of the Virasoro
algebra. This would give the triple
\s$({\NH\hs{0.03cm}'}^{\rm R}_0\s,\s
Q^{\rm cs\s\pm}_0|_{{\NH\hs{0.02cm}'}^{\rm R}_0}\s,
\s\CA'_0)\s$, \s where
\qq
{\NH\hs{0.02cm}'}^{\rm NS,R}_0\s=\s
\{\ |\phi\rangle\in\NH^{\rm NS,R}\ \s |\s \ Q^{\rm cs\s\pm}_n\s
|\phi\rangle=0=L^{\rm cs\s\pm}_n\s
|\phi\rangle\s,\s\ {\rm for}\ \s
n>0\s ,
\ \s (L^{\rm cs\s+}_0-L^{\rm cs\s-}_0)\s|\phi\rangle=0\ \}
\hs{0.4cm}
\qqq
and \s$\CA'_0\s$ is generated by sandwiching
the primary fields
\s$V^{\rm cs}_{|\phi\rangle}(1)\s$ corresponding
to vectors \s$|\phi\rangle\in{\NH\hs{0.02cm}'}^{\rm NS}_0\s$
between orthogonal projections onto
\s${\NH\hs{0.02cm}'}^{\rm R}_0\s$. \s
The above construction of the effective target
is  consistent for any N=1 super-CFT.
\vs 0.3cm

An interesting open problem is to find natural
connections associated with the coset targets
and to study their flatness
properties. In the quantum mechanical case
the coset geometry corresponds to a dilatonic deformation
of Riemannian geometry. The presence of the
dilaton changes the perturbative
conditions for the conformality of the sigma
models \cite{Callan} so that the presence of non-trivial
Ricci curvature in the coset target should not be
surprising. One may expect a similar deformation
to occur for the non-commutative effective targets of the
coset models of field theory defined above.
\vskip 0.8cm

\nsection{\hspace{-.7cm}.\ \ N=2 CFT and mirror symmetry}
\vs 0.2cm
\no{\bf{5.1.\ \ \s N=2 SUSY coset models}}
\vs 0.4cm

In \cite{KazSuz}\cite{KazSuz1},
Kazama and Suzuki pointed out that there is a class of
SUSY \s$G/H\s$
coset models which possess extended N=2
superconformal symmetry.
The latter requires two series of operators \s$G^\pm_n\s$
and \s$\bar G^\pm_n\s$ extending the Virasoro algebras
\s$(L^\pm_n)\s$ and a commuting pair of \s$u(1)\s$ current
algebras \s$(j^\pm_n)\s$ with the commutation relations
\qq
[j^\pm_n\s,\s j^\pm_m]\s\ \ \ &=&\ {_1\over^3}\s
c\s n\s\delta_{n+m,0}\ ,\cr
[j^\pm_n\s,\s G^\pm_m]\ \ \s&=&\ G^\pm_{n+m}\ ,\cr
[j^\pm_n\s,\s \bar G^\pm_m]\ \s\ &=&\ -\bar G^\pm_{n+m}\ ,\cr
[L^\pm_n\s,\s G^\pm_m]\ \ \hs{-0.03cm}&=
&\ ({_n\over^2}-m)\s G^\pm_{n+m}\ ,\cr
[L^\pm_n\s,\s \bar G^\pm_m]\hs{-0.03cm}\ \ &=
&\ ({_n\over^2}-m)\s \bar G^\pm_{n+m}\ ,\cr
\hs{-0.05cm}\{G^\pm_n\s,\s \bar G^\pm_m\}\s
&=&\ 2\s L^{\pm}_{n+m}\s+
\s(n-m)j^\pm_{n+m}\s+\s{_1\over^3}
\s c\s(n^2-{_1\over^4})\s\delta_{n+m,0}\ .
\qqq
The operators $G^\pm_n\s$ commute
with \s$L^\mp_m\s$ and \s$j^\mp_m\s$
and anticommute with \s$G^\mp_m\s$ and \s$\bar G^\mp_m\s$.
The N=2 superconformal symmetry implies the N=1 one:
we may set \s$Q^\pm_n={1\over{\sqrt{2}}}(G^\pm_n+
\bar G^\pm_n)\s$. \s
One may realize the N=2 algebra in SUSY coset models if
\s$G/H\s$ is a homogeneous K\"{a}hler manifold, in
particular, if  \s$G/H\s$ is a hermitian symmetric space.
In the latter situation, one may decompose
\qq
({\bf h}^\perp)^\NC={\bf t}\oplus\bar{\bf t}\ ,
\qqq
where \s${\bf t}\s$ and \s$\bar{\bf t}\s$ are complex
conjugate abelian Lie subalgebras of \s${\bf g}\s$,
preserved (\s mod \s${\bf h}^\NC\s$) \s by the adjoint action
of \s$\bf h\s$, \s
isotropic for \s$\tr\s$ and s.t.
\s$[{\bf t},\bar{\bf t}]\subset{\bf h}\s$. We shall
choose a basis \s$(t^\alpha,t^{\bar\alpha})\s$
of \s$({\bf h}^\perp)^\NC\s$ with the property that
\s$t^{\bar\alpha}
=\overline{t^\alpha}\s$ and that \s$\tr\s\hs{0.03cm} t^\alpha
t^{\bar\beta}={1\over 2}\s\delta^{\alpha\beta}\s$.
Then
\qq
G^\pm_n&=&{_2\over^{\sqrt{k}}}\s\sum\limits_m\psi^\alpha_m
J^{\bar\alpha}_{n-m}\ ,\cr
\bar G^\pm_n&=&{_2\over^{\sqrt{k}}}\s\sum\limits_m
\psi^{\bar\alpha}_m J^{\alpha}_{n-m}\ ,\cr
j^\pm_n\ &=&{_{k-g^\vee}\over^k}\sum\limits_m
:\psi^{\alpha\pm}_m\psi^{\bar\alpha\pm}_{n-m}:\s
-\s{_{2i}\over^k}\s
f^{{\rm a}\alpha\bar\alpha}J^{{\rm a}\pm}_n\ .
\qqq
\vs 0.3cm

The simplest hermitian symmetric space
is \s$P\NC^1=SU(2)/U(1)\s$.
It gives the {\bf minimal} N=2 {\bf series} of the SUSY coset
theories with central charges \s${3(k-2)\over k}\s$.
The space of states of these models is
a finite sesquilinear
combination of irreducible unitary representations
of the N=2 superconformal algebra
(with \s$c<3\s$)\s \cite{Boucher}.
Another example of a hermitian symmetric space
is provided by the Grassmannian
\s$SU(n+m)/SU(n)\times SU(m)\times U(1)\s$.
It leads to a series of \s N=2 coset models with
central charges \s${3nm(k-n-m)\over k}\s$. The complete
list of (compact) hermitian symmetric spaces is short \cite{Helg}
and contains, besides the above examples, still
\s$SO(2n)/SO(n)\times SO(2)\s$, \s$SO(2n)/SU(n)\times U(1)\s$,
\s$SP(n)/SU(n)\times U(1)\s$, \s$E_6/SO(10)\times U(1)\s$ and
\s$E_7/E_6\times U(1)\s$.
\vs 0.3cm

We may identify the effective target geometry,
\s$(\NH^{\rm R}_{\s0}\s,\s Q^{\rm cs\s\pm}_0
|_{{\NH}^{\rm R}_{\s0}}\s,\s\CA_0)\s$, \s of \s
N=2 SUSY coset
models in the same way as for the N=1 ones.
Alternatively, we may pose
\qq
{\NH\hs{0.03cm}'}^{\rm NS,R}\hs{-0.43cm}_0
=\{\ |\phi\rangle\in
\NH^{\rm NS,R}\ \s |\ \ L^\pm_n,\s G^\pm_n,\s\bar G^\pm_n,\s
j^\pm_n|\phi\rangle
\s\smash{\mathop{=}\limits_{n>0}}\s0\ ,
\ \s (L^+_0-L^-_0)|\phi\rangle=0\ \}
\qqq
and define the small algebra \s$\CA'_0\s$ to be the one
generated by
the primary fields
\s$V_{|\phi\rangle}(1)\s$, \s$|\phi\rangle\in
{\NH\hs{0.03cm}'}^{\rm NS}\hs{-0.25cm}_0\s$,
multiplied by the orthogonal projection onto
\s${\NH\hs{0.03cm}'}^{\rm R}\hs{-0.25cm}_0\s$
from both sides\footnote{\s it is also natural to require
here that \s$j^+_0|\phi\rangle=j^-_0|\phi\rangle=0\s$}.
The latter construction may be done
for any N=2 conformal model.
Notice that
\s$\NH^{\rm R}_{\s0}\s$ (\s${\NH\hs{0.03cm}'}^{\rm R}
\hs{-0.25cm}_0\s$) is preserved by the
operators \s$G^\pm_0\s,\ \s\bar G^\pm_0\s$ and \s$j^\pm_0\s$
which, when restricted to \s$\NH^{\rm R}_0\s$
(\s${\NH\hs{0.03cm}'}^{\rm R}\hs{-0.25cm}_0\s$)\s,
\s satisfy the algebra:
\qq
&(G^\pm_0)^2=0=(\bar G^\pm_0)^2\ ,\hspace{1cm} \ \ \ \ \ \
\{G^+_0\s,\s\bar G^+_0\}=\{G^-_0\s,\s\bar G^-_0\}\ ,&\cr
&\ \ \ [j_0^\pm\s,\s G^\pm_0]=G^\pm_0\ ,\ \
[j_0^\pm\s,\s \bar G^\pm_0]=-\bar G^\pm_0\ ,\ \
(G^\pm_0)^*=\bar G^\pm_0\ ,\ \ (j^\pm_0)^*=j^\pm_0&\ .
\label{ALG}
\qqq
The operators with superscript $+$ (anti)commute
with the ones with superscript $-$. This is the same algebra
as the one satisfied by operators
\s$G^\pm\equiv\Gamma^{\alpha\pm}\nabla_\alpha\ ,\s\
\bar G^\pm\equiv\Gamma^{\bar\alpha\pm}\nabla_{\bar\alpha}\s$ and
\s$j^\pm\equiv{1\over 4}\gamma_{\alpha\bar\beta}
\s[\Gamma^{\bar\beta\pm},
\s\Gamma^{\alpha\pm}]\s$ acting on sections of
the bundle \s$S\otimes S\s$ (or \s$S\otimes S\otimes \NC^2\s$) \s
over a K\"{a}hler manifold \s$M\s$ with the K\"{a}hler metric
\s$\gamma_{\alpha\bar\beta}\s dz^\alpha d{\overline{z^{\beta}}}\s$
and the metric connection \s$\nabla\s$.  Equivalently,
we may consider the operators \s$\da\s,\ \de^*,\ \da^*,\ \de\s$
on \s$L^2(\bigwedge M)\s$ and interpret \s$\pm j^\pm\s$
as counting the degrees
in \s$\bigwedge^{p,q}M\s$.
Since \s$(G^\pm_0)^2=0=(\bar G^\pm_0)^2\s$, \s we may consider
the cohomology of any of those operators. In fact, all
these cohomologies coincide and may be identified with the subspace
of \s$\NH^{\rm R}\s$
annihilated by \s$G^\pm_0\s$ and \s$\bar G^\pm_0\s$.
By analogy to the situation in the K\"{a}hler geometry,
we shall call this subspace ``harmonic'' and we shall denote it
by \s$\NH^{\rm harm}\s$.
It is easy to see that \s$\NH^{\rm harm}
\subset{\NH\hs{0.03cm}'}^{\rm R}\hs{-0.07cm}_0\s$.
It coincides with the space of Ramond sector
ground states corresponding to the eigenvalue
\s${c\over{24}}\s$ of \s$L^\pm_0\s$.
\vs 0.3cm

In the classical case, the space of harmonic forms
on a K\"{a}hler manifold \s$M\s$ may be identified with the
de Rham cohomology ring of \s$M\s$. Does the ring
structure have a counterpart in \s$\NH^{\rm harm}\s$?
In N=2 theory (with integral \s$j_0+\bar j_0\s$ charges),
there exists a unitary transformation \s$U\s$ from
\s$\NH^{\rm R}\s$ to \s$\NH^{\rm NS}\s$
(the spectral flow \cite{SeiSchw}) \s s.t.
\qq
\hbox to 3.1cm{$U^{-1}(L^\pm_n\mp\hf j^\pm_n)\s U$\hfill}
&=&\s\s L^\pm_n-{_c\over^{24}}
\s\delta_{n,0}\ ,\cr
\hbox to 3.1cm{$U^{-1} j^\pm_n\s U$\hfill}
&=&\s\s j^\pm_n\pm{_c\over^6}
\s\delta_{n,0}\ ,\cr
\hbox to 3.1cm{$U^{-1} G^\pm_{n\mp{1/2}}U$\hfill}
&=&\s\s G^\pm_n\ ,\cr
\hbox to 3.1cm{$U^{-1}\bar G^\pm_{n\pm1/2}U$\hfill}
&=&\s\s\bar G^\pm_n\ .
\qqq
$U\s$ maps \s$\NH^{\rm harm}\s$ onto the so called chiral-chiral
primary subspace \s$\NH^{\rm(c,c)}\s$ of \s$\NH^{\rm NS}\s$
annihilated by \s$G^\pm_{\mp1/2}\s$ and
\s$\bar G^\pm_{\pm1/2}\s$. The primary fields
\s$V_{|\phi\rangle}(z^\pm)\s$ corresponding
to \s$|\phi\rangle\in\NH^{\rm(c,c)}\s$ have non-singular
operator product, i.e. they may be multiplied
point-wise. Moreover, their product is the vertex operator
of another (possibly vanishing)
 state in \s$\NH^{\rm(c,c)}\s$. \s This way
 \s$\NH^{\rm (c,c)}\s$ becomes a graded-\break commutative
``chiral-chiral'' ring \cite{LerVaWar}.
By the action of \s$U\s$, \s the ring structure may be carried
over to \s$\NH^{\rm harm}\s$.
We may then think about the data
\qq
(\NH^{\rm R}_0\s,\s G^+_0\s,\s
\bar G^-_0\s,\s j^\pm_0\s,\s
\CA_0)\ \ \ {\rm or}\ \ \ ({\NH\hs{0.03cm}'}^{\rm R}
\hs{-0.07cm}_0\s,\s
G^+_0\s,\s\bar G^-_0\s,\s j^\pm_0\s,\s
\CA'_0)\ \ \ \ \ \ {\rm and}\ \ \ \ \ \NH^{\rm (c,c)}\ \s
\label{N=2data}
\qqq
as representing the effective target geometry and cohomology
of the N=2 CFT, a deformation of the K\"{a}hler geometry
\s$(L^2(\bigwedge M),\s \da,\s\de,\s j^\pm,\s C^\infty(M))\s$
and of its de Rham cohomology.
\vs 0.3cm

It has been shown in \cite{Gep}\cite{Gep1}
that for certain orbifolds of
the tensor products of minimal N=2 models
which have integral \s$j^\pm_0\s$
charges, the rings \s$\NH^{\rm (c,c)}\s$ are
deformations of the de Rham cohomology rings of certain
Calabi-Yau manifolds\footnote{\s i.e. K\"{a}hler manifolds
with SU(3) holonomy and, consequently, Ricci flat}.
In particular, the dimensions of the spaces of harmonic
states with \s$j^\pm_0\s$ charges \s$p-{c\over 6}
\hs{0.02cm},\s{c\over 6}-q\s$
are equal to the Hodge numbers\footnote{\s i.e. to the dimensions
of the spaces of harmonic $(p\hs{0.02cm},\hs{0.02cm}q)$-forms}
\s$h_M^{p\hs{0.02cm},\hs{0.02cm}q}\s$ of
a Calabi-Yau space \s$M\s$ of complex dimension \s$d=
{c\over 3}\s)\s$. \s
One may also
argue directly (see E. Witten's contribution to \cite{Yau})
that, for sigma models
with Calabi-Yau targets, the chiral-chiral ring is the
cohomology ring of the manifold deformed by
instanton effects\footnote{\s instantons of the sigma model with
K\"{a}hler target \s$M\s$ are complex curves in \s$M\s$}.
The deformation disappears in the semi-classical limit sending
the radius of the target space to infinity.
One should expect that, in an appropriate classical limit,
also the effective
target geometry of the N=2 models, not only
their cohomology, becomes that of
Calabi-Yau manifolds or of their
torsion or/and dilatonic versions, see M. Ro$\check{\rm c}$ek's
contribution to \cite{Yau}. This issue deserves further study.
It should be remarked, that a complex version
of non-commutative geometry, appropriate for study
of data (\ref{N=2data}), still remains to be developed.
\vs 0.4cm
\vs 0.2cm
\no{\bf{5.2.\ \ \s Mirror targets}}
\vs 0.4cm

The N=2 super-conformal algebra does not change if
we reverse the sign of the \s$u(1)\s$ current \s$j_n\s$
interchanging at the same time \s$G_n\s$ and \s$\bar G_n\s$.
Constructing the effective target geometry and cohomology
in a given N=2 model with integral \s$j^\pm_0\s$ charges
after a replacement \s$j^-_n\mapsto
-j^-_n\s, \s G^-_n\leftarrow\hs{-0.3cm}\rightarrow
\bar G^-_n\s$, one obtains
different objects: mirror effective target
and the chiral-antichiral ring:
\qq
(\NH^{\rm R}_0\s,\s G^+_0\s,\s
G^-_0\s,\s \pm j^\pm_0\s,\s
\CA_0)\ \ \ {\rm or}\ \ \ ({\NH\hs{0.03cm}'}^{\rm R}
\hs{-0.25cm}_0\s,\s
G^+_0\s,\s G^-_0\s,\s\pm j^\pm_0\s,\s
\CA'_0)\ \ \ \ \ \ {\rm and}\ \ \ \ \ \NH^{\rm (c,a)}\ .
\qqq
It has been checked that
in some cases \s$\NH^{\rm (c,a)}\s$ is a
deformed cohomology ring
of a different Calabi-Yau manifold \s$\tilde M\s$ with
the Hodge numbers
\qq
h_{\tilde M}^{p\hs{0.02cm},\hs{0.02cm}q}
=h_M^{p\hs{0.02cm},\hs{0.02cm}d-q}\ .
\nonumber
\qqq
Besides, the chiral-chiral ring \s$\NH^{\rm (c,c)}\s$
should then coincide with the undeformed Dolbeault
cohomology ring with values in the exterior algebra
of the holomorphic tangent bundle of \s$\tilde M\s$.
The latter is usually easier to compute.
This way,
one may extract non-trivial
information about instanton numbers for a Calabi-Yau
manifold \s$M\s$, \s usually hard to obtain directly,
from an easy calculation of the Dolbeault cohomology
on the mirror image \s$\tilde M\s$ of \s$M\s$
\cite{AspMorr}\cite{BershCOV}.
\s The mirror symmetry \s$M\leftarrow\hs{-0.2cm}
\rightarrow \tilde M\s$ should extend
from special pairs, where it was directly
verified, to the moduli space of Calabi-Yau spaces,
with the interchange of the role of
moduli of complex structures (counted by
\s$h^{d-1\hs{0.02cm},\hs{0.02cm}1}
=h^{1\hs{0.02cm},\hs{0.02cm}d-1}\s$)
and of K\"{a}hler structures (corresponding to
\s$h^{1\hs{0.02cm},\hs{0.02cm}1}\s$)\s.
\vs 0.3cm

The N=2 conformal models are difficult to control
out of special points but the cohomological
information given by the chiral-chiral
or chiral-antichiral rings is contained in
the topological field theories, easier to solve,
obtained by coupling
the \s$u(1)\s$ currents of the conformal models to
the spin connection \cite{Witten}\cite{EguYang}.
The chiral-chiral and chiral-antichiral rings
reduce to the de Rham cohomology rings of Calabi-Yau
spaces \s$M\s$ and \s$\tilde M\s$
in different semi-classical limits. Although
the geometry of the complete CFT is more difficult
to control then its cohomology, the construction which
associates to a given N=2 theory a mirror pair of effective
(non-commutative) targets
\qq
({\NH\hs{0.03cm}'}^{\rm R}_0\s,\s
G^+_0\s,\s\bar G^-_0\s,\s j^\pm_0\s,\s
\CA'_0)\ \ \ {\rm and}\ \ \
({\NH\hs{0.03cm}'}^{\rm R}_0\s,\s
G^+_0\s,\s G^-_0\s,\s\pm j^\pm_0\s,\s
\CA'_0)
\qqq
is essentially tautological:
non-commutative geometry in its complex version
should provide a natural setup for the mirror symmetry.
\vs 0.3cm

The simplest example of a mirror pair of Calabi-Yau
spaces is obtained by taking the complex three-dimensional
smooth projective variety \s$M\s$ defined by the equation
\qq
z_1^5+z_2^5+z_3^5+z_4^5+z_5^5=0
\qqq
in \s$P\NC^4\s$ \cite{GreenPless}.
Its canonical bundle has vanishing
first Chern class and, by the theorem of Yau
proving Calabi's conjecture, it admits a metric
with \s$SU(3)\s$ holonomy (unique up to normalization).
The corresponding CFT is a projected version of
the product of five copies of
the minimal
\s$k=5\s$ N=2 models. The mirror image \s$\tilde M\s$
of \s$M\s$ is the orbifold\footnote{\s more
exactly, after the resolution of its singularities}
of \s$M\s$ under the action of the \s$\NZ^3_5\s$ group
generated by
\qq
(z_1,z_2,z_3,z_4,z_5)&\mapsto&(z_1,\rho z_2,\rho^2 z_3,\rho^3 z_4,
\rho^4 z_5)\ ,\cr
(z_1,z_2,z_3,z_4,z_5)&\mapsto&(z_1,\rho z_2,\rho z_3,\rho^4 z_4,
\rho^4 z_5)\ ,\cr
(z_1,z_2,z_3,z_4,z_5)&\mapsto&(z_1,z_2,z_3,\rho z_4,
\rho^4 z_5)\ ,
\qqq
where \s$\rho\s$ is a fifth root of \s$1\s$.
The relevant Hodge numbers are \s$h_M^{1,1}=h_{\tilde M}^{2,1}=1\s$,
\s$h_M^{2,1}=h_{\tilde M}^{1,1}=101\s$.
Many other examples of mirror pairs of Calabi-Yau
spaces were explicitly identified, see
\cite{GreenPless}\cite{CLSch}\cite{Yau},
and the subject is under intensive study
\cite{Schimm}\cite{CandDerP}\cite{Candelas}
\cite{HKTY}\cite{BershCOV}\cite{AGM}.
\vs 1.2cm

\end{document}